\documentclass[fleqn,10pt]{wlscirep}

\usepackage[utf8]{inputenc}
\usepackage[T1]{fontenc} 
\usepackage{amsmath}
\usepackage{mathtools}
\usepackage{mathpazo} 

\title{Dynamic core periphery structure of information sharing networks in entorhinal cortex and hippocampus}

\author[1,2]{Nicola Pedreschi}
\author[2]{Christophe Bernard}
\author[2]{Wesley Clawson}
\author[2]{Pascale Quilichini}
\author[1,3]{Alain Barrat}
\author[2]{Demian Battaglia}

\affil[1]{Aix Marseille Univ, Universit\'e de Toulon, CNRS, CPT, Turing Center for Living Systems, Marseille, France}
\affil[2]{Aix Marseille Univ, Inserm, INS, Institut de Neurosciences des Syst\`emes, Marseille, France}
\affil[3]{Tokyo Tech World Research Hub Initiative (WRHI), Institute of Innovative Research, Tokyo Institute of Technology.
Japan}
\date{}
\providecommand{\keywords}[1]{\textbf{\textit{Keywords}} #1}

\begin{document}


\begin{abstract}
Neural computation is associated with the emergence, reconfiguration and dissolution of cell assemblies in the context of varying oscillatory states.
Here, we describe the complex spatio-temporal dynamics of cell assemblies through temporal network formalism.  We use a sliding window approach to extract sequences of networks of information sharing among single units in hippocampus and enthorinal cortex during anesthesia and study how global and node-wise functional connectivity properties evolve along time and as a function of changing global brain state (theta vs slow-wave oscillations). First, we find that information sharing networks display, at any time, a  core-periphery structure 
in which an integrated core of more tightly functionally interconnected units link to more loosely connected network leaves.
However the units participating to the core or to the periphery substantially change across time-windows, with units entering and  
leaving the core in a smooth way. Second, we find that discrete network states can be defined 
on top of this continuously ongoing liquid core-periphery reorganization. Switching between network states results in a more abrupt modification of the units belonging to the core and is only loosely linked to transitions between global oscillatory states. Third, we characterize different styles of temporal connectivity that cells can exhibit within each state of the sharing network. While inhibitory cells tend to be central, we show that, otherwise, anatomical localization only poorly influences the patterns of temporal connectivity of the different cells. Furthermore, cells can change temporal connectivity style when the network changes state. Altogether, 
these findings reveal that the sharing of information mediated by the intrinsic dynamics of hippocampal and enthorinal cortex cell assemblies have a rich spatiotemporal structure, which 
could not have been identified by more conventional time- or state-averaged analyses of functional connectivity.

\end{abstract}

\maketitle

\section{Introduction}

Since its early definitions \cite{Hebb1949, Abeles1982}, the notion of \textit{cell assembly}, loosely defined as a group of neurons with coordinated firing within a local or distributed circuit, has been associated to information processing. According to a widely accepted view (see e.g. \cite{Varela:2001bs}), neuronal representations and, more generally, computations are constructed via the dynamic integration of the information conveyed by the spiking activity of different cells. The recruitment of a cell assembly goes thus well beyond the mere co-activation of an ensemble of cells frequently firing together. It corresponds indeed to the instantiation of an actual transient functional network allowing information to be shared between the involved neurons and ultimately fed into novel informational constructs suitable for further processing \cite{Buzsaki2010}.

In this sense, it appears quite natural to describe the flexible materialization, transformation and dissolution of groups of information-exchanging neurons as a dynamic network whose nodes and edges evolve along time, reflecting the recruitment (or the dismissal) 
of neurons into (or out of) the current integrated assembly. Nevertheless, cell assemblies at the level of neuronal microcircuits ---and, particularly, in the hippocampal formation, involved in spatial navigation and episodic memory \cite{Buzsaki:2013ji}--- have been most often characterized in terms of sets of nodes frequently co-activating in time \cite{Mao2001, Miller:2014eq} or repeatedly activating in sequences \cite{Ikegaya:2004ds, Malvache:2016dd}. Such ``static'' catalogues of patterns of firing partially fail at highlighting that the temporally coordinated firing of nodes 
gives rise to a dynamics of functional links \cite{Aertsen1989, Bonifazi:2009in, Clawson2019}, i.e., 
to a \textit{temporal network} \cite{Holme:2012jo, holme2015modern,holme:2017}.

The temporal network framework has recently emerged in order to take into account that for many systems, a static network representation 
is only a first approximation that hides very important properties. This has been made possible by the availability of temporally resolved
data in communication and social networks in particular  \cite{Barabasi2005Nature,Cattuto2010PLOS,Karsai2011PhysRevE,Saramaki:2015}:
studies of these data have uncovered features such as broad distributions of contact or inter-contact times (burstiness) between individuals
\cite{Barabasi2005Nature,Cattuto2010PLOS},  
multiple temporal and structural scales  \cite{Saramaki:2015,Darst,Laurent:2015}, and a rich array of intrinsically dynamical 
structures that could not be unveiled within a static network framework  \cite{Kovanen:2011,Gauvin2014,galimberti2018mining,Kobayashi:2019}.
Taking into account temporality has moreover been shown to have a strong impact in processes taking place on networks, 
in particular the propagation of diseases or of information \cite{Karsai2011PhysRevE,Ribeiro2013SciRep,holme:2019}.
In the neuroscience field, emphasis have been recently put on the need to upgrade ``connectomics'' into ``chronnectomics'', to disentangle temporal variability from inter-subject and inter-cohort differences and thus achieve a superior biomarking performance \cite{hutchison2013, calhoun2014}. However a majority of dynamic network studies have been so far considering large-scale brain-wide networks of interregional connectivity –--see, e.g. \cite{Thompson:2017} for a study of link burstiness--- and only fewer have addressed the dynamics of information sharing networks at the level of micro-circuits, often in vitro or in silico \cite{Stetter:2012be, Orlandi:2013bm, Poli2015} and even more rarely in vivo \cite{Clawson2019}.

Here, we propose to fully embrace a temporal network perspective when describing dynamic information sharing within and between cell assemblies. Concretely, we analyze high-density electrophysiological recordings in the hippocampus and medial entorhinal cortex of the rat, allowing to follow in parallel the spiking activity of several tens of single units across different laminar locations in different brain regions. We focus on recordings performed during anesthesia, guaranteeing long and stable recordings of intrinsic cell assembly dynamics over several hours \cite{Quilichini:2010bt}. Indeed, even during anesthesia, the spatiotemporal complexity of firing is not suppressed but a wide repertoire of co-firing ensembles and associated information processing modes can be found \cite{Clawson2019}. Furthermore, with the used anesthesia protocol, we observe a characteristic stochastic alternation between two global brain oscillatory states, respectively dominated by Slow Oscillations (SO) and Theta (THE) oscillations, reminiscent of a slowed-down version of natural sleep, with its alternation between non-REM and REM epochs (see \textit{Methods}). Such experimental condition is therefore particularly suitable to probe the dependence of cell assembly dynamics on the currently active global oscillatory state \cite{Clawson2019}, which is expected to be a major modulator of information processing in the hippocampus formation \cite{Klausberger2003, Buzsaki:2013ji} and cortical circuits in general \cite{Varela:2001bs, Gilbert:2007hb}. Using a sliding window approach and estimating functional connectivity via an analysis of the pairwise mutual information between the spike trains of distinct single units ---following \cite{Clawson2019}---, we extract temporal networks of information sharing and we develop methods to investigate  
how their connectivity properties evolve in time.

At the level of whole network organization, we pay attention to whether connectivity structure changes continuously ---as in the case of many social or communication networks \cite{Cattuto2010PLOS,Miritello:2013,Saramaki:2015}--- or rather undergoes switching between different discrete network states,
---as recently observed for instance 
in an animal social network \cite{gelardi:2019}---, possibly in relation with transitions between global oscillatory states. We find that switching between \textit{discrete network states} does spontaneously occur during anesthesia. Remarkably, 
we identify a multiplicity of states of connectivity between single neurons, with a rich switching dynamics ongoing even in absence of a 
change in the global oscillatory state. The sharing network connectivity, however, is never frozen, but keeps fluctuating even within each of the network states. More specifically, at any time, the instantaneous information sharing network displays 
a core-periphery organization \cite{Rossa,Kojaku:2018}, in which a limited number of neurons 
form a  tightly mutually connected core, while a majority of other neurons are more peripheral. 
Individual neurons flexibly modulate along time their degree of integration within the sharing network and may ``float'' 
between the core and the periphery, transiently leaving or getting engaged into the core, giving rise to what we 
call a \textit{liquid core-periphery} architecture.

At the level of single nodes, the neighborhood of each neuron is  changing smoothly within network states and more abruptly  
across a state switching. New connections can be formed or old connections disappear and the number of neighbors can vary in time. 
Most importantly, 
neurons who have at the time aggregated level similar static connectivities, having connections for  a comparable overall amount of time,
can strongly differ in their temporal connectivity profile. For instance, some neurons may form links 
that remain active for a limited amount of time but in an uninterrupted way. Other neurons may instead 
repeatedly connect and disconnect to others, sharing information in an intermittent and sporadic fashion. 
We thus define for each neuron its specific temporal connectivity profile, which summarizes its dynamic 
patterns of attachment in the evolving core-periphery sharing network.
We then use an unsupervised classification approach to identify \textit{temporal connectivity style archetypes} 
and show that neurons can adopt different styles in different network states, possibly associated to variations of their role in information processing (see \textit{Discussion}).

Going beyond averaged network analyses with suitable time-resolved metrics allows thus an unprecedented precision in characterizing the functional organization of information sharing. Cell assemblies are not anymore seen as rigidly defined groups of cells but as  
dynamic networks, restlessly exchanging flows of information between core and periphery and continuously modifying their extent and reach toward cells in different anatomical locations. In other words, the adoption of a temporal network framework makes it 
possible to witness and seize the inner life of cell assemblies while it unfolds and gives rise to emergent computations.


\section{Results}\label{results}

\subsection{Information sharing dynamics can be described as a temporal network}

\begin{figure}[htb]
\centerline{\includegraphics[width=.8\textwidth]{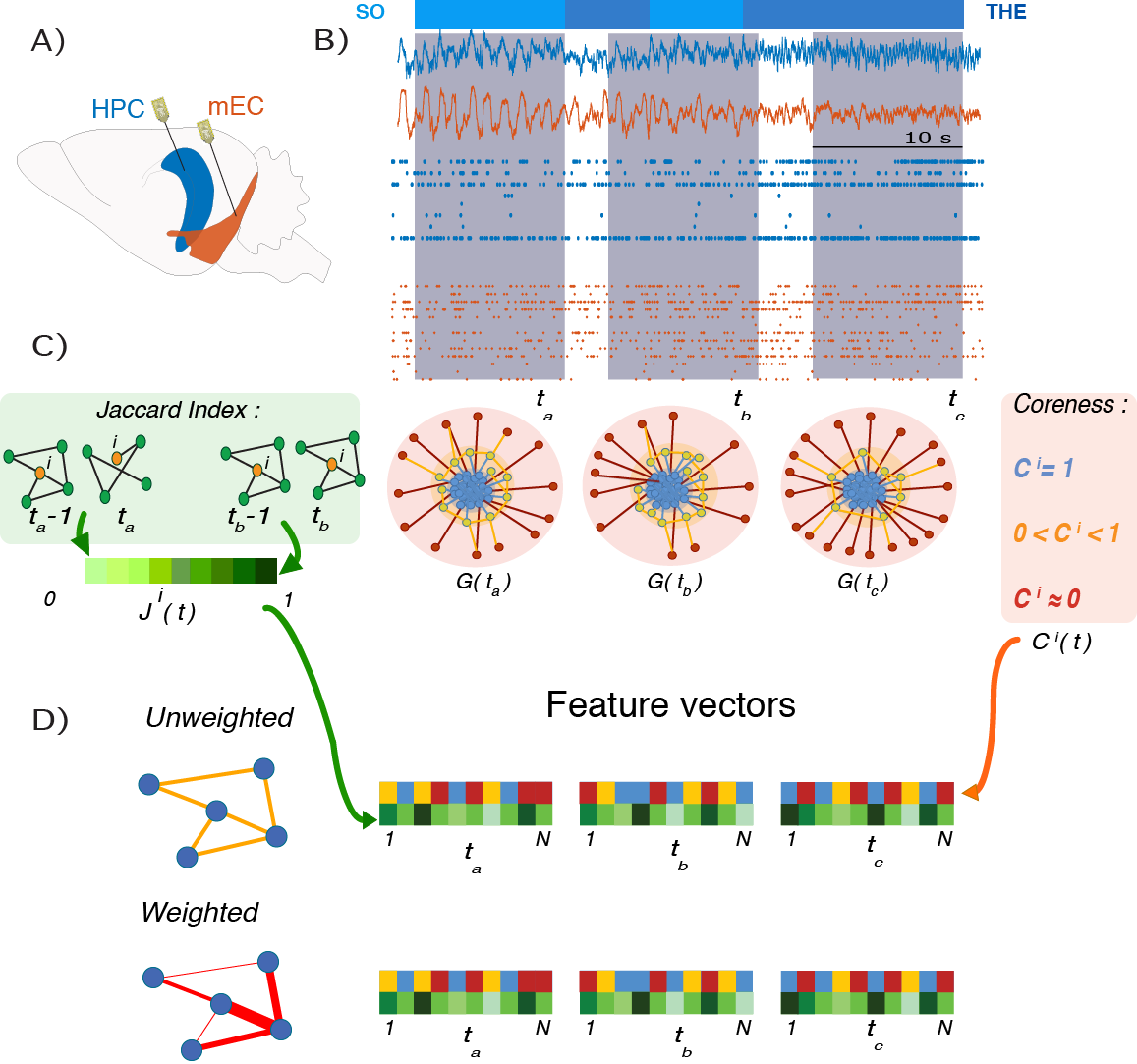}}
\caption{\textbf{Recordings and feature vectors.}
 \textbf{A)} Approximate recording locations in mEC (orange) and CA1 (blue) during anesthesia.
 \textbf{B)} An example of LFP signal recorded by the channels in CA1 and in mEC; below the LFPs are two examples of single unit activity  
 from the same recording. The intervals $t_a$, $t_b$ and $t_c$ are examples of time windows in which the corresponding network time-frames
 are constructed. The horizontal bar above the recordings represents the Global Oscillations states: in light blue the \emph{Slow Oscillations} and in darker blue the \emph{Theta Oscillations}. 
 \textbf{C)} Illustrative sketches of 
 the two features computed for each node at each time-frame: on the left, in the green box, we illustrate the \emph{Jaccard index}, which quantifies the overlap between two sets. 
 Here we consider the local Jaccard indices between the neighborhoods of each node $i\in[1,N]$ in successive time frames (e.g., $J^i(t_a)$
 is the Jaccard index for node $i$ between times $t_a-1$ and $t_a$): these quantities
 carries information on how little or how radically the neighborhoods of $i$ changes across time
 and we call them \emph{liquidity}. 
 The center and right panels are a schematic representation of the core-periphery organization of the network and the intuitive explanation of the \emph{coreness} $C$ values of the nodes: \emph{core} (blue) nodes have high values of coreness $C\sim 1$, whereas peripheral (red) nodes have low values, $C\sim 0$ ; the yellow area surrounding the core is the \emph{core-skin}, i.e., nodes whose coreness values are higher than $0$ and lower than $1$. When a node is \emph{disconnected} from the rest of the network (when it has no neighbours), its coreness is exactly zero. 
 \textbf{D)} At each timeframe, we obtain 
 feature vectors of the coreness and
 liquidity values of all neurons $i\in[1,N]$,
 schematized by the two colourful vectors  
 (yellow, blue and red for coreness and green for liquidity). The analysis were performed both
 for unweighted networks, in which links are either present or absent, and for weighted networks, in which the links are characterized by a weight
 given by the amount of shared information between the nodes. For weighted networks, liquidity is measured by the cosine similarity instead of the
 Jaccard index (see \textit{Methods}), and a weighted coreness is computed for each node.} 
\label{fig:fig1}
\end{figure}

\begin{figure}[htb]
\centerline{\includegraphics[width=\textwidth]{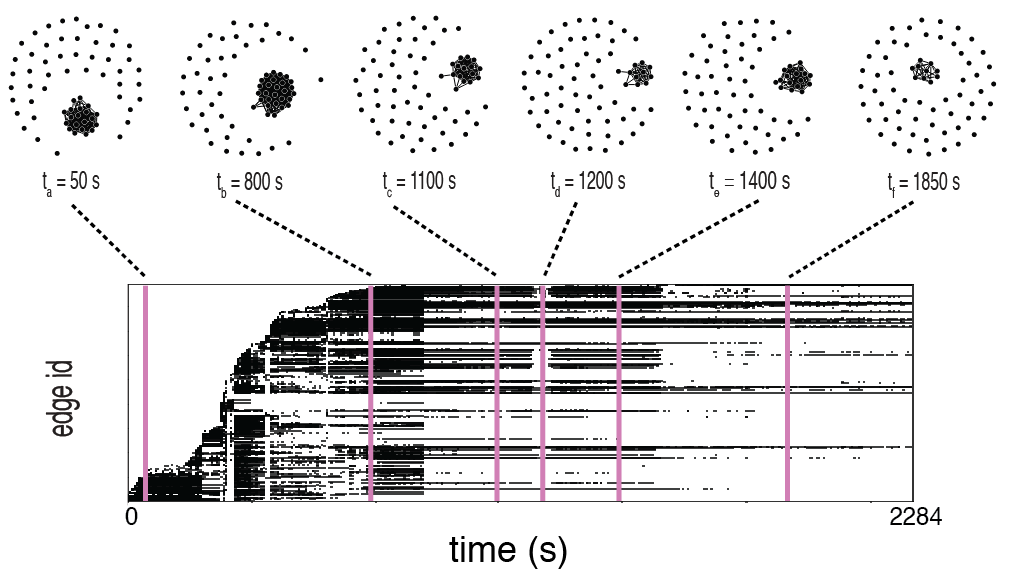}}
\caption{\textbf{Temporal network visualization.}
 In this figure we represent the evolution of the information sharing temporal network computed for one of the recordings, 
 using the visualization toolbox TACOMA
 (\texttt{http://tacoma.benmaier.org/about.html}). 
 The six networks on the top plots represent six snapshots of the network's evolution at the corresponding times: each black dot represents a neuron whose activity has been recorded at the corresponding time. 
 The bottom panel displays the temporal network's \emph{edge activity plot}: each row of the plot represents the activity of one 
 edge of the network, i.e., a black dot on a row means that the corresponding edge on the vertical axis is active (a non-zero element of the network's adjacency matrix) at the corresponding time on the horizontal axis. The edges are ordered on the vertical axis by increasing value of their
 \emph{first activation time}.} 
\label{fig:fig0}
\end{figure}

Single unit recordings of neuronal activity were acquired simultaneously from the CA1 region in the Hippocampus and in the mEC (Figures \ref{fig:fig1}.A
and \ref{fig:S1}) for 16 rats under anaesthesia (18 recording for 16 rats). Following \cite{Clawson2019}, we 
constructed time-resolved weighted networks of functional connectivity, adopting a sliding window approach.
Within each $10$s long time-window, we took connection weights (functional links) between pairs of neurons (network nodes) to be proportional to the amount of \emph{shared information} \cite{Kirst:2016jt} between their firing rates (see \textit{Methods}). We then slide the time window by $1$s, in order to achieve a $90\%$ overlap between consecutive windows. In Figure \ref{fig:fig1}.B, we represent the temporal network construction procedure. 
Based on the data segments in each of three windows centered at times $t_a$, $t_b$ and $t_c$, we extract a $N \times N$ matrix for
each time window, where $N$ is the number of neurons and in which the element $(i,j)$ corresponds to the shared information between nodes $i$ and $j$. Each such matrix, in network terms, is interpreted as the \emph{adjacency matrix} 
of a weighted graph $G$ of $N$ nodes. 
Even if the used functional connectivity metric is in principle pseudo-directed (because of the presence of a time-lag between putative sender and receiver node, see \textit{Methods}), 
we found that asymmetries between reciprocal connections were very small (see \textit{Methods}), 
especially for the stronger connections, and chose therefore to symmetrize the adjacency matrix for most analyses.
This procedure thus maps each multi-channel 
recording of length $T$ seconds to a time series of $T$ network representations, obtaining finally a 
temporal network of information sharing among neurons, formed by the temporal succession of these $T$ network snapshots. 
Cartoon representations of the temporal network snapshots 
$G(t_a)$,  $G(t_b)$ and $G(t_c)$ in the three highlighted time windows are shown in Figure \ref{fig:fig1}.C. Some
actual network frames of a specific recording, together with a diagram describing emergence and disappearance of
links (\emph{edge activity plot}) can be seen in Figure \ref{fig:fig0}.

On the top of Figure \ref{fig:fig1}.B we also present the characteristic switching between global oscillatory states observed in our recordings. Analysis of the local field potentials recorded simultaneously to single unit activity allowed identifying a spontaneous stochastic-like alternation between epochs belonging to a first SO global state (light blue color) spectrally dominated by $<1$ Hz 
oscillations and epochs in a second THE global state, characterized by the presence of high power in a 4-8 Hz spectral band. In the following we will relate the temporal network reconfiguration dynamics to these global oscillatory state transitions.

We carried on our study by analyzing in parallel the evolution of the weighted temporal network structure and 
of the corresponding unweighted temporal network. To this aim we defined  for each time window
a binarized version of the network snapshot, whose adjacency matrix is only composed of zeros and ones: 
the adjacency matrix element is $0$ (no link is present in the unweighted network) 
when two nodes are not connected (the shared information between them is zero), 
and it is equal to $1$ when they are (the shared information is nonzero). 
By comparing weighted and unweighted analyses we could assess whether the
network changes involve actual evolutions of the structures of the links or rather correspond only to weight modulations on a stable 
link structure. To quantify the dynamics of the network, we focused on two main aspects ---translated into corresponding temporal network features in both weighted and unweighted versions---, aiming at answering  
two different questions. First: are the connections of the network stable in time, or rapidly changing? 
Second: does the network have a clear and specific structural organization, and if so, is it persistent in time or unstable and only transient?

In order to answer the first question, we quantified, for each neuron $i$, how much its neighborhood changed 
between successive time windows \cite{Miritello:2013,Valdano:2015,gelardi:2019,STEHLE2013604}. 
To this aim we computed for each $i$ and at each time $t$ the 
 \emph{cosine similarity} $\Theta^i(t)$
 between the neighborhoods
 of $i$ (the subgraphs composed only by the edges involving $i$) 
 at time $t-1$ and at time $t$. 
 To analyze the unweighted temporal networks, we instead used the \emph{Jaccard index} $J^i(t)$ between these successive neighborhoods
 (see \textit{Methods} for precise definitions). 
 Values of these quantities close or equal to $1$ suggest that the node has not changed neighbors in successive time windows:
hence its neighborhood shows low \emph{liquidity} (elsewhere, it would be said that the node shows high ``loyalty'' \cite{Valdano:2015}). 
On the contrary, values  close or equal to $0$ mean that the neuron has completely changed neighbors between subsequent times: its neighborhood is highly liquid. 
 At each time $t$, the set of cosine similarity values $\Theta^i(t),\ i\in[1,N]$ and
 the Jaccard index values $J^i(t),\ i\in[1,N]$ 
 (for the unweighted case) 
 form the time-dependent feature vectors $\mathbf{\Theta}(t)$ and
 $\mathbf{J}(t)$, each of dimension $N$
(Figure \ref{fig:fig1}.D).

In order to answer the second question and probe for the presence of specific network architectures, we considered the
\emph{core-periphery} organization of the graph. This way of characterizing the information sharing network
snapshots was suggested to us by the visual inspection of their 
spatial embeddings, some of which are represented in Figure \ref{fig:fig0}. 
We thus computed the \emph{coreness coefficient} $C^i(t)$ of each node $i$ in each snapshot $t$, using
the definition of  coreness introduced by \cite{Rossa} (see \textit{Methods}). 
In  a static, unweighted, undirected network, the coreness $C^i$ of node $i\in[1,N]$ is a real number between $0$ and $1$, 
interpreted as follows: when $C^i\sim 1$ the node belongs to the \emph{core} of the network, i.e., a set of tightly connected nodes; 
when $C^i\simeq 0$ the node belongs instead to the network's \emph{periphery}, i.e., is only loosely linked to the rest of the network; 
if $C^i=0$ the node is actually disconnected, i.e., its degree (number of neighbors) is $k^i=0$. 
We thus obtain a time dependent vector 
$\mathbf{C}(t)$ of dimension $N$ by computing at each time $t$ the coreness
$C^i(t)$ for each node $i\in[1,N]$ in the network of time-window $t$.
The computation of coreness coefficient can also be performed for 
weighted networks \cite{Rossa}, possibly yielding, however, values larger than $1$. Therefore, we normalize the whole time series of vectors of weighted coreness coefficients by the maximum 
observed value in order to obtain a time series of vectors $\{\mathbf{C}_w(t)|t\in[0,T]\}$ 
(with $\mathbf{C}_w(t) = \{C^i_w(t), i\in[1,N]\}$)
with the same
range of values for the unweighted and weighted coreness features
(Figure \ref{fig:fig1}.D).

\subsection{Information sharing networks have a soft core-periphery architecture}

In order to investigate the core-periphery organization of the information sharing networks, we looked at the distributions of the instantaneous coreness values $C^i(t)$ over all neurons and time-frames. 
A weighted coreness distribution from a representative recording is shown in Figure \ref{fig:fig2}.A 
{(see Supplementary Figure \ref{fig:S2} for equivalent unweighted coreness analyses)}. 
Figure \ref{fig:fig2}.B moreover displays  
instantaneous distributions
of the weighted coreness, for the same recording
and for several time frames.
We found that, within each time frame, a majority of neurons had low coreness values, i.e., they were peripheral nodes 
in the instantaneous sharing network (red color in the cartoons of Figure \ref{fig:fig2}.A),
while fewer neurons had high coreness values. Interestingly, in most recordings there was not a sharp separation between core and periphery. 
On the contrary, we generically observed the presence of a smooth distribution spanning all possible coreness values. 
The transition between core and periphery was thus smooth, without gaps but with neurons displaying gradually less 
tight links with the core, without yet being fully peripheral. We note that such smooth distributions are actually encountered in many systems
\cite{Rossa}, a strict distinction between a very central core and a very loose periphery being only a schematic idealized vision and real
networks displaying typically hierarchies of scales and local connectivities.

The analyses of Figure \ref{fig:fig2}.A and \ref{fig:fig2}.B indicate that at any time-frame the sharing network has a soft core-periphery architecture, 
but does not inform us about how individual neurons evolve in time within this architecture. In order to follow dynamic changes in the coreness of individual neurons, we studied the time-evolution of this feature for each neuron of each recording. 
In Figure \ref{fig:fig2}.C we plot the coreness $C_w^i(t)$ vs time, for each node $i\in[1,N]$ of a representative recording. 
The two highlighted lines in the figure represent the coreness evolution of two particular nodes. In light green, we show the instantaneous
coreness of the node with maximum average coreness $\langle C_w^i(t)\rangle_T$ (averaged over the recording length $T$). 
The figure shows clearly that this neuron's instantaneous coreness is always large: the corresponding neuron is persistently part of the network's core throughout the whole recording.  This contrasts with the purple line, which displays the instantaneous coreness of the neuron
with largest coreness standard deviation ($\sigma( \langle C_w^i(t) \rangle_T)$): the curve fluctuates 
from high to low coreness values, indicating that the corresponding neuron switches several times between 
central core positions in the network and more peripheral ones. The contrast between these two behaviors is highlighted
in the cartoon at the bottom of Figure  \ref{fig:fig2}.C.

The continuous range of observed
instantaneous coreness values and the
fluctuations in individual coreness values 
indicate that the set of most central neurons changes in time. We thus 
examined whether some regions
were contributing more than others to this core. To this aim, we define the core, at each time-frame, as the set of neurons whose
instantaneous coreness lies above the $95$th percentile of the distribution (in
histograms such as those in Figure \ref{fig:fig2}.B).
 In Figure \ref{fig:fig2}.D we then plot the \emph{core filling factors} of the CA1 and mEC layers (top and center plots, respectively). 
 We define the core filling factor of each region as the percentage of the overall number of neurons of the recording located in that region that 
 belong to the core. We plot the 
 time-evolution core filling factors 
 separately for neurons located in different hippocampal CA1 layers (light and dark blue lines, top panel) and for neurons in different medial entorhinal cortex (mEC) layers (red, orange and yellow lines, center panel). 
 The figure illustrates thet the core-filling factors vary substantially along time. In the example shown here
 (corresponding to the same recording as in Figure \ref{fig:fig2}.C), the core-filling factor of CA1 Stratum Pyramidale (SP) 
 neurons belonging to the core increases from $\sim 2\%$ to near $7\%$ during the recording. 

The results of Figure \ref{fig:fig2}.D indicate that the
core is not restricted to neurons belonging to a specific region, but is generally composed of both neurons 
belonging to EC and neurons belonging to CA1. We remind indeed that our networks are networks of functional connectivity and do not have to reflect necessarily the underlying anatomical connectivity (for which it would be unlikely that our recordings pick up mono-synaptically connected cells between different regions). However, the participation of CA1 and EC neurons to the core is changing through time and, as a result, the core is
 sometimes ``more on the EC side'' or ``more on the CA1 side'' (see lower cartoons in Figure \ref{fig:fig2}.D). 
 To visualize the relative fractions of core neurons belonging to the two different regions, we 
computed and show in the bottom panel  of Figure \ref{fig:fig2}.D the normalized \emph{core filling regional fractions}: the fraction of core nodes belonging either to EC (orange) or CA1 (blue) -- the sum of these two fractions adding up to $1$.
The orange and blue bands change thickness along time, reflecting in this recording 
a progressive shift from a low to a 
higher involvement of CA1 neurons in the core. This variation of the multi-regional core composition may reflect changes in the 
way the different regions control information integrative processes (see \textit{Discussion}). 

\begin{figure}[htb]
\centerline{\includegraphics[width=.8\textwidth]{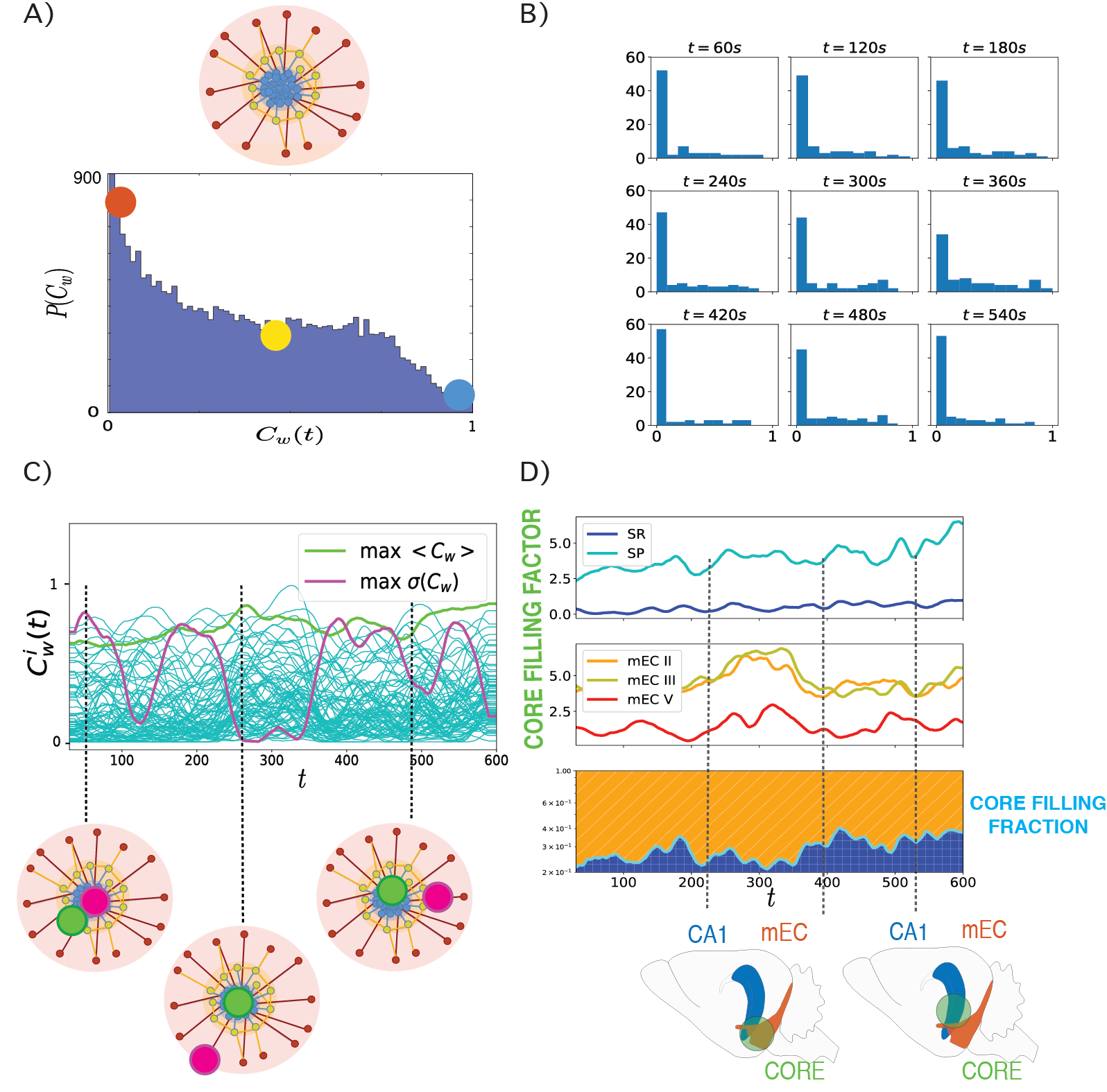}}
\caption{\textbf{Dynamic Core-Periphery structure.} 
\textbf{A)} Top: Cartoon representing the \emph{core-periphery} organization of the network with \emph{core} nodes colored in blue and \emph{peripheral} nodes in red; Bottom: histogram
of the values of the instantaneous weighted coreness  $C^i(t)$ of each neuron at each time of a recording: the most frequent values of $C_w^i(t)$ are close to $0$ (red dot), hence coreness values typical of \emph{peripheral} nodes, whereas 
 coreness values of core nodes, $C^i(t)\sim 1$ are  rare (blue dot).  
 \textbf{B)} Histograms of instantaneous weighted coreness values of the neurons in several individual
 time-frames (every $60$ seconds). The different histograms show that the picture of a majority of neurons with low coreness value and few neurons with high coreness holds at all times. 
 The core periphery organization of the information sharing network is thus persistent in time. 
 \textbf{C)} Plot of the temporal evolution of the weighted coreness $C_w^i(t)$ of each node $i$ in a specific recording: the red line highlights the evolution of the coreness of the node with highest average coreness $\langle C_w(t)\rangle_{max}$ (time average over the whole recording); the orange line corresponds to the node whose coreness fluctuates the most (node of maximum variance of $C_w^i(t)$). 
 We show below this plot a cartoon of the core-periphery organization of the network, where we illustrate how the former node 
  steadily belongs to the core, while the latter node evolves several times from the core to the periphery and back. 
  \textbf{D)} Plots of the \emph{core-filling factor} (top and center panels) and of the \emph{core-filling fraction} (bottom) of the different layers of \emph{hippocampus} (HPC) and medial \emph{entorhinal cortex} (mEC). The core-filling factor  of each layer is the percentage of nodes of each layer that are located
  in the core (nodes above the $95th$ percentile of coreness values). In the bottom panel, 
  the green line represents the separation 
  between the fraction of core nodes located either in the mEC layers (orange) or in the HPC (blue): the \emph{core filling fraction}. The cartoon
  below the core-filling fraction plot illustrates how the neurons that belong to the core at different times can be located in different anatomical structures.} 
\label{fig:fig2}
\end{figure}

\subsection{Network states} 

\begin{figure}[htb]
\centerline{\includegraphics[width=\textwidth]{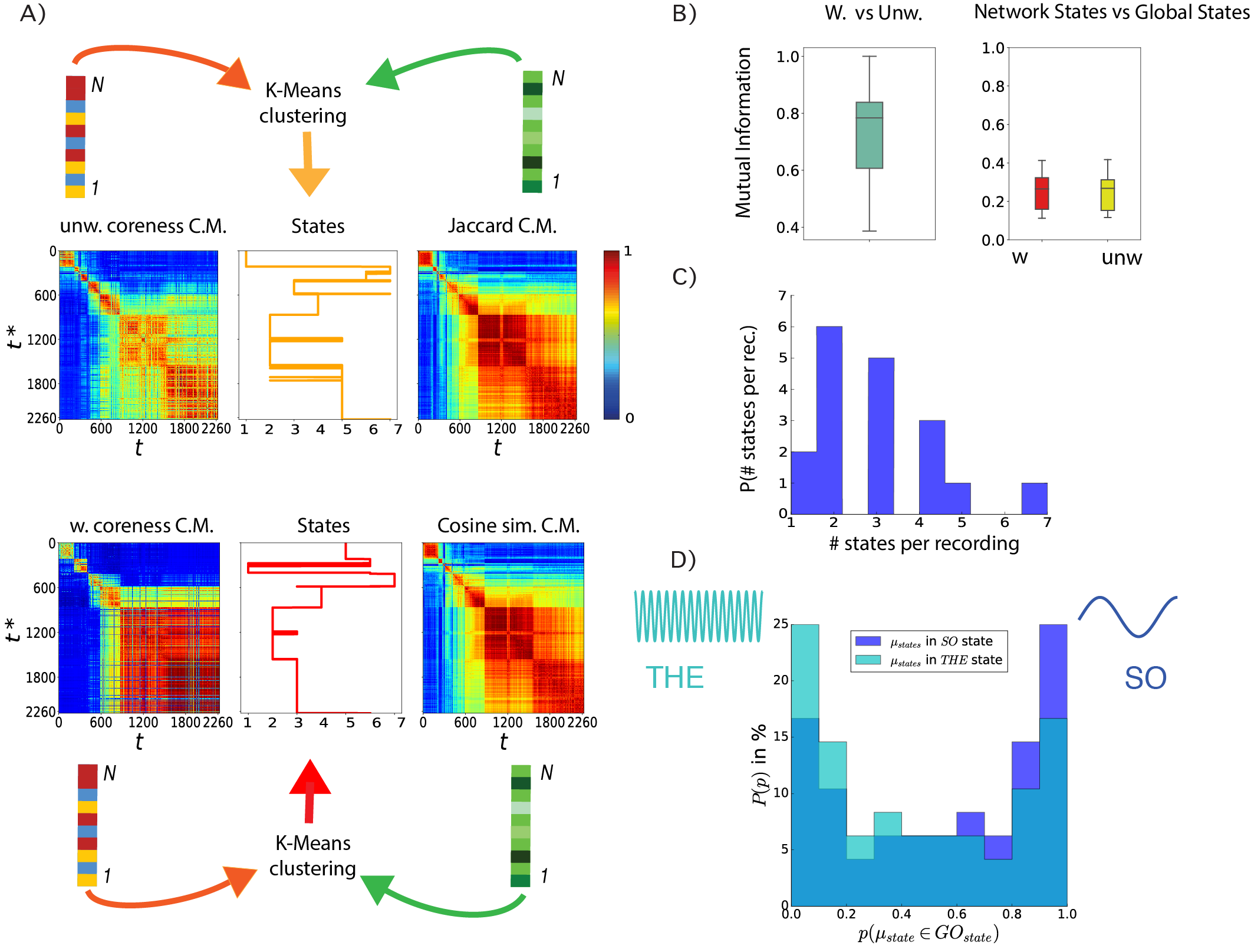}}
\caption{\textbf{Liquidity and coreness network-states.} 
\textbf{A)} For each recording, the feature vectors of liquidity 
and coreness described in Figure
\ref{fig:fig1}D are combined at each time,
obtaining time-dependent vectors of dimension $2N$:
$\{\mathbf{J}(t), \mathbf{C}(t)\}$ in the
unweighted case, 
$\{\mathbf{\Theta}(t), \mathbf{C}_w(t)\}$ for
the weighted case. K-means clustering of these
vectors yields for each recording a sequence of \emph{network states} visited by the network (centre panels). We show on both sides of the network state sequences the correlation matrices (C.M.) 
of the various feature vectors
(unweighted coreness and Jaccard, weighted
coreness and cosine similarity):
each element $(t,t^*)$ of a matrix is the Pearson correlation coefficient between the corresponding 
feature vectors computed at times $t$ and $t^*$.
For instance, the cosine similarity C.M. 
has as element $(t,t^*)$ the Pearson correlation 
between the list of values of the 
cosine similarity (weighted liquidity) of nodes at times $t$ and $t^*$, $\{\Theta^i(t), i=1,\dots,N\}$ and $\{\Theta^i(t^*),i=1,\dots N \}$. 
\textbf{B)} Left panel: Distribution of the values of the mutual information between the weighted and unweighted network states spectra of all recordings: 
the mutual information between the two spectra is high, showing that the two spectra are highly correlated for all recordings. 
Right panel: distribution of the values of mutual information between the weighted network states spectrum and the global oscillating states spectrum (red boxplot) for all recordings
(yellow boxplot: same for the unweighted network states spectra). 
\textbf{C)} Statistics of the 
number of network states in the different recordings. In many recordings this number is larger than $2$, and it can reach values as large as $7$.
\textbf{D)} 
Histograms showing
 the percentage of \emph{states} having 
 probability $p$ to occur during a specific \emph{Global Oscillations state}, \emph{slow} (blue) or \emph{theta} (light blue). Both
 histograms are bi-modal, hence the majority of \emph{states} are strongly \emph{global-state}-specific; the blue histogram shows an abundance of states ($\sim 23\%$) most likely to occur during the \emph{slow} global oscillations.}
\label{fig:fig3}
\end{figure}

As previously mentioned and summarized in Figure \ref{fig:fig1}.C, for each recording and each time window $t$, we computed for each neuron $i\in[1,N]$
several temporal network properties, tracking notably the ``liquidity'' of its neighborhood (Jaccard index and cosine similarity) 
and its position within the core-periphery architecture (weighted and unweighted instantaneous coreness values). 
To investigate how these properties change dynamically at the global network level, we computed for each of these four quantities
the correlation between their values at different times, obtaining four correlation matrices of size $T \times T$. For instance, 
the element $(t,t')$ of the unweighted liquidity correlation matrix is given by the Pearson correlation between the $N$ values
of the Jaccard coefficient computed at $t$ $\{J^i(t)$, $i\in[1,N]\}$ and the $N$ values computed at $t'$
$\{J^i(t')$, $i\in[1,N]\}$ (see \textit{Methods} for definition). In Figure \ref{fig:fig3}.A, we show these 
four correlation matrices for a representative recording, 
two for the unweighted features (above, Jaccard index and unweighted coreness), and two for the weighted ones (below,
cosine similarity and weighted coreness). 

The block-wise structure of these correlation matrices suggests the existence of  
epochs in time where neurons' feature values are strongly correlated (red blocks on the diagonal). 
In the case of Figure \ref{fig:fig3}.A, mostly diagonal blocks are observed, with low correlation values outside the blocks,
meaning that the network configurations are similar during each epoch but very different in different epochs.
In other cases, we sometimes observe as well off-diagonal blocks, 
indicating that the network might return to a configuration close to one previously observed 
(we show an example of this behavior
in Supplementary Figure \ref{fig:S3}, 
as well as an example in which only
one epoch is observed).
Each block on the diagonal (epoch in which the node properties are strongly correlated) can be interpreted as 
a  network connectivity configuration associated to specific liquidity and coreness assignments of 
the various neurons. 
We call \emph{network states} these configurations. 

To quantitatively extract such discrete network states, we use the time-series of the feature vectors 
$\mathbf{\Theta}(t)$, $\mathbf{J}(t)$, $\mathbf{C}_w(t)$ and $\mathbf{C}(t)$. 
We concatenate these vectors
two by two at each time, obtaining two $2N$-dimensional feature vectors: the first one contains at each time
the values of the unweighted liquidity and coreness of all nodes ($\{\mathbf{J}(t), \mathbf{C}(t)\}$),
and the second one contains the corresponding weighted values
($\{\mathbf{\Theta}(t), \mathbf{C}_w(t)\}$). 
We then perform in each case (weighted and unweighted) an unsupervised clustering of 
these $T$ $2N$-dimensional feature vectors. 
As a result of this clustering procedure, as shown in  Figure \ref{fig:fig3}.A, we obtain a sequence of states 
(temporal clusters of the feature vectors) 
that the network finds itself in at different times (yellow state spectrum for the unweighted case, red for the weighted case).
We also observe that these states are not a mere artificial construction. Our procedure would segment the temporal network into states even if discretely separated clusters did not exist, but we explicitly checked that, for all recordings but two, clustering was meaningful: Supplementary Figure \ref{fig:S4}.A shows indeed that clustering quality was in a large majority of cases well above chance level (see \textit{Methods}).

We compared the network states spectra found for the weighted and unweighted case by computing the mutual information between the two sequences of states for each recording, normalized by the largest entropy among the entropies of the two distinct sequences. 
Such relative mutual information is bounded in the unit interval and quantifies the fraction of information that 
a state sequence carries about the other (reaching the unit value when the two state sequences are identical, and being zero 
if the two sequences are statistically independent). We compute this quantity for each recording and display the distribution of values obtained
as a light green boxplot on the left of Figure \ref{fig:fig3}.B. 
This boxplot shows that the mutual information values between the weighted and unweighted network states sequences of a recording
are concentrated around a median approaching $0.8$. Therefore, the spectra of network states extracted by the weighted and unweighted analyses are generally matching well, indicating the robustness of their extraction procedure. 
Most importantly, the high degree of matching between weighted and unweighted analyses confirms that network state changes correspond to actual connectivity re-organizations (as revealed by unweighted analyses) and not just to weight modulations on top of a fixed  
connectivity.

As previously discussed, the system undergoes switching between two possible global brain states during the anesthesia recordings:
these global states are associated to different oscillatory patterns, dominated by either Theta (THE) or slow (SO) oscillations. 
We studied therefore what is the relation occurring between changes in the network state and global oscillatory state switching. 
To this aim, we computed the normalized mutual information between network state sequences (weighted or unweighted) and global state sequences. The distributions of the values obtained are shown as boxplots 
on the right of Figure \ref{fig:fig3}.B, for both weighted (red) and unweighted (yellow) network states sequences. 
In both cases, we detect positive, although low, values of the relative mutual information with global oscillatory states, 
with a median value close to $\sim 0.3$. 
This indicates that some coordination between global oscillatory state and network state switching exists but that oscillatory state switching does not well explain network state switching. 
A very simple reason for this poor correlation is that, while there are just two main global oscillatory states (see however \textit{Discussion}), 
the number of network states is not a priori limited. In fact, the statistics of the number of network states in the different recordings,  
shown in Figure \ref{fig:fig3}.C, indicates that in 
many recordings we could extract at 
 least three network states and sometimes up to seven. Therefore, network state switching can occur within each oscillatory global state.

Nevertheless, it is possible that each given network state would tend to occur mostly within one specific global oscillatory state. 
To check whether this is the case, we computed for each network state the 
fraction of times that this state occurred 
during THE or SO epochs. We show in Figure \ref{fig:fig3}.D the histograms of these time fractions, measured over the set of all network states. The light blue histogram corresponds to the fractions of time a network state manifested itself during the THE state (the dark blue histogram gives the same information but for the SO state). Both histograms are markedly bimodal, indicating that 
a majority of states occur a large fraction of times during either the THE or the SO states, but not in both. In other words, 
network states are to a large degree oscillatory state specific. Therefore, the global oscillatory states do not fully determine the 
observed coreness and liquidity configurations (there may be several network states for each of the oscillatory states) 
but most network states 
can be observed only during one specific global
oscillatory state and not during the other.

\subsection{Connectivity profiles and styles}

\begin{figure}[htb]
\centerline{\includegraphics[width=.8\textwidth]{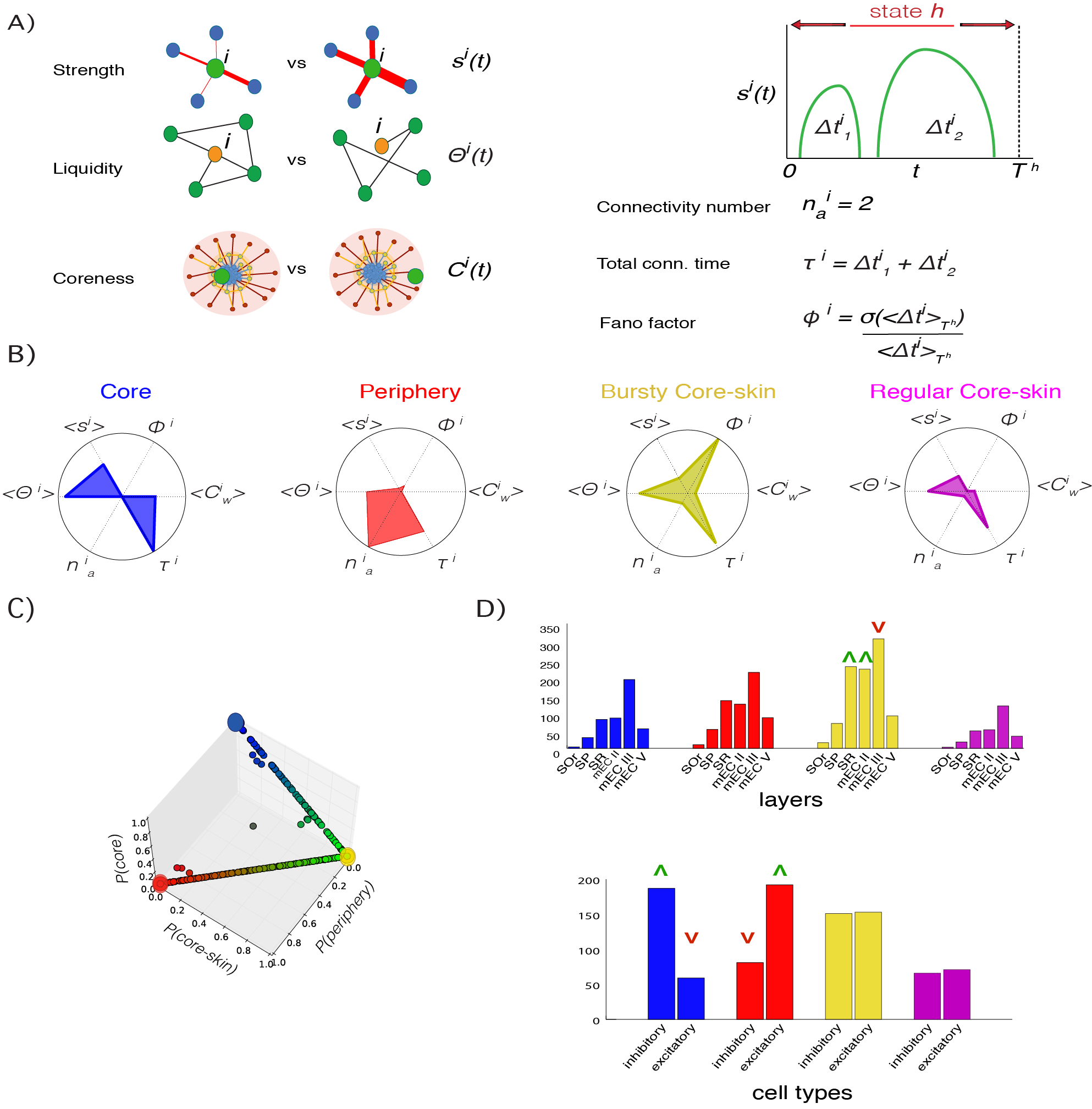}}
\caption{ \textbf{Connectivity profiles.} 
\textbf{A)} Cartoons illustrating the features: the \emph{strength} $s_i(t)$ of a node $i$ at time $t$
measures the global importance of node $i$'s connections (the cartoon shows a comparison of low vs high values of $s^i(t)$); its liquidity, hence the cosine similarity (Jaccard index in the unweighted case) between successive neighborhoods (the cartoon illustrates a change between the neighborhood of $i$ at successive times); its coreness; 
the connectivity number $n^i_a$ and total connectivity time $\tau^i$ of node $i$ in a state: $\tau^i$ represents the number of time frames in which neuron $i$ is connected to at least one other neuron, and $n^i_a$ is 
the number of times that $i$ switches from being disconnected to being connected to at least one other neuron; the Fano factor quantifies the fluctuations
of the durations of the $n^i_a$ periods in which $i$ is connected.  
\textbf{B)} The connectivity profile
of each neuron $i$ in a state $h$ is a 6-dimensional vector whose components are its \emph{state-averaged} features: average coreness
$\langle C^i_w\rangle_h$, average liquidity
$\langle \Theta^i\rangle_h$, average strength
$\langle s^i\rangle_h$, connectivity number $n_a^i$, total connectivity time $\tau^i$, Fano factor $\Phi^i$.
Clustering of all the connectivity profiles yields 4 \emph{connectivity styles} that we represent
here as radar plots. We identify four main styles: core (blue), periphery (red), bursty core-skin (yellow) and regular core-skin (magenta), with different distinctive values of the different features composing the connectivity profile. Shown here are radar plots for the connectivity profile of archetypal cells in each of the connectivity styles (see Results for details).
  \textbf{C)} 3D plot of the connectivity styles space: the three axis are the probabilities 
  of the connectivity profile of a node in a network state to belong to the periphery, core-skin or core, respectively. 
  \textbf{D)} Histograms of the layer location
  (top plot)
  of the neurons exhibiting each connectivity style (identified by the same color as panel B), and histograms of cell type (inhibitory and excitatory, bottom plot) populations per connectivity style. A green upward arrow means that the connectivity style of the corresponding color is statistically over-represented in the corresponding layer or cell type, whereas a red downward arrow means that the connectivity style is under-represented in that layer or cell type.}
\label{fig:fig4}
\end{figure}

To refine our analysis, we now investigate and characterize the temporal network properties at the level of single neurons 
within each of the detected network states. In order to do so, we computed, for each node and in each state, a set of dynamical features 
averaged over all time frames assigned to the specifically considered state. We focus here on the weighted features, since 
the weighted and unweighted analysis provide similar results.
The state-specific \emph{connectivity profile} of a given neuron $i$ in a given state
included first:
\begin{itemize}
\item its state-averaged weighted coreness value;
\item its state-averaged cosine similarity value.
\end{itemize}
Note that analogous time-resolved features were already used for network state extraction, but that we consider
here state-averaged values.
We also computed four additional network state specific features, defined as follows 
for each node $i\in[1,N]$ in a state $h$ spanning the set of times $T^h$ 
(see explanatory cartoons in Figure \ref{fig:fig4}.A):
\begin{itemize}
    \item the state-averaged strength $\langle s^i(t) \rangle_{t \in T^h}  \equiv\langle \sum_j w_{ij}(t)\rangle_{t \in T^h}$, 
    hence the state-averaged total instantaneous weight of the connections of $i$;
    \item the activation number $n^i_a$: it is the number of times that the strength of node $i$ changes from $0$ to a non-zero value
    within the state, hence it gives the number of time that $i$ changes its connectivity from being isolated to 
    being connected to at least one another neuron;
     \item the total connectivity time $\tau^i$: it is the number of time frames within $T_h$ in which $i$ is connected to at least one another neuron;
    \item the Fano factor $\Phi^i\equiv\frac{\sigma(\langle \Delta t^i\rangle_{h})}{\langle \Delta t^i\rangle_{h}}$: 
    each of the  $n^i_a$ periods in which $i$ is connected to at least another neuron has a certain duration $\Delta t^i$
    (the sum of these durations is $\tau^i$), and  $\Phi^i$, the ratio of the variance and the average of  the 
     different connectivity durations of $i$ over the state $h$, 
    quantifies whether these durations are of similar value or very diverse.
\end{itemize}{}
Each neuron's temporal properties within a given state are thus summarized by the 
values of these overall six features, which define the neuron's state-specific connectivity profile as a six-dimensional vector.
 Normalizing all the features to have values between $0$ and $1$, 
 connectivity profiles can be visually represented as radar plots (Figure \ref{fig:fig4}.B) in which the value of each feature is plotted on the corresponding radial axis.

After computing the connectivity profile of each node, in each network state and in each recording, 
we performed an unsupervised clustering (using K-means clustering) over all these six-dimensional 
connectivity profiles in order to identify categories of these profiles, which we call \emph{connectivity styles}. 
With this approach we uncovered the existence of four general connectivity styles that a neuron can manifest within a network state:
\begin{itemize}
    \item \emph{Core style (or ``streamers'')}, a class of nodes of high average coreness, average strength, average cosine-similarity, total connectivity time and low activation number $n_a$ and Fano factor $\Phi$: overall, a class of rather central neurons with numerous stable connections that are persistently connected within a state (blue representative polygone in Figure \ref{fig:fig4}.B). A behavior similar to that of a speaker of an assembly, continuously conveying information to the same, and many, people: a \emph{``streamer''} of information;
    \item \emph{Peripheral style (or ``callers'')}, nodes with high activation number and total connectivity time but 
    low strength, Fano factor and coreness: a class of peripheral nodes that are periodically connected in numerous events of
     similar connectivity durations and low weights within a state, whose connections are not completely liquid 
     nor completely stable (red representative 
     polygone in Figure \ref{fig:fig4}.B). A behavior similar to a customer or guest regularly 
     making short calls to trusted core members to be updated on the latest news: aka, a ``caller''.  
    \item \emph{Bursty} and \emph{Regular core-skin style (or ``free-lancer helpers'' and ``staff helpers'')}, 
    two classes of connectivity profiles both characterized by nodes with intermediate coreness and strength and high values of cosine similarity 
    and total connectivity time, whose difference lies in the values of the Fano factor $\Phi$: the former (yellow representative polygone in Figure \ref{fig:fig4}.B) displaying high values of $\Phi$ can therefore be interpreted as a class of nodes that have stable connections, that are active for a long time yet with highly varying connectivity times; the latter (purple representative polygone in Figure \ref{fig:fig4}.B) is characterized instead by low values of the Fano factor, hence the connectivity durations of these nodes do not fluctuate much. 
    The behaviors of these neurons can be assimilated to the one of external experts assisting core staff in a company, either with regular work schedules (the regular core-skin neurons could then be seen as \emph{``staff helpers''}) or sporadically and irregularly  (the bursty core-sking neurons could then be seen as \emph{``free-lancer  helpers''}).
\end{itemize}{}
Note that we also identified (and subsequently discarded) an additional ``Junk'' cluster with relatively fewer elements ($9.7\%$ of the total number of connectivity profiles computed for all neurons in all recordings) and small 
values of all features.  
We thus removed these cases, as usual in unsupervised clustering applications \cite{Forsyth2018}, 
to  better discriminate the remaining ``interesting'' classes listed above. 

Overall each connectivity profile (one for each neuron in each possible network state in the associated recording) was categorized as belonging to one of the above connectivity styles, according to the output cluster label assigned by the unsupervised clustering algorithm. 
However, a substantial diversity of connectivity profiles subsists within each of the clusters. 
We therefore considered as well a soft classification scheme, which quantifies the degree of relation of each individual 
connectivity profile with the tendencies identified by each of the different connectivity style clusters. 
Concretely, we trained a machine learning classifier to receive as input a connectivity profile and predict the connectivity style assigned to it by 
this unsupervised clustering. 
In this way, after training, the classifier assigned to each connectivity profile a four-dimensional vector
whose elements represented the probabilities of belonging to any one of the four possible connectivity styles 
(see \textit{Methods} for details). In Figure \ref{fig:fig4}.C, 
each connectivity profile is represented as a dot with as
coordinates the soft classification labels produced by the classifier, i.e.,  
the probabilities that each given connectivity profile belongs to the periphery, 
core-skin (summing the probabilities for the bursty and regular subtypes) or core connectivity styles. 
This plot reveals, on the one hand, the existence of a gap between core (blue) and periphery (red) connectivity profiles: 
in other words, connectivity profiles that are likely to be classified as of the ``streamer'' type are very unlikely to be classified as 
being of the ``caller'' type, stressing the radical difference between these two connectivity styles. 
On the other hand, both the core and periphery connectivity styles display some mixing with the core-skin style, as 
made clear by the almost continuous paths of connectivity profiles from the core (blue) to the core-skin (yellow) 
and from the core-skin to the periphery (red). This means that there is a continuum spectrum of connectivity profiles interpolating between ``streamers'' and ``helpers'' on one side and ``helpers'' and ``callers'' on the other. 
The polygones shown in Figure \ref{fig:fig4}.B are on the contrary \emph{archetypal} (in the sense introduced by \cite{Battaglia:2013kk}). 
These archetype profiles manifest in an extreme manner the tendencies inherent to their connectivity style. This is reflected by the fact that they lie at the vertices of the bounded soft membership space represented in Figure \ref{fig:fig4}.C. They display thus strong similarity to just one connectivity style, which they epitomize even better than the centroids of the associated connectivity style cluster, having near zero chance 
of being misclassified
(cluster centroids are shown for comparison in 
Supplementary Figure \ref{fig:S4}.B). 

We finally checked whether the different connectivity styles were adopted more or less frequently by neurons in specific anatomical locations or of specific types (excitatory or inhibitory). 
In Figure \ref{fig:fig4}.D, we plot the number of connectivity profiles assigned to each style 
(colors as in Figure \ref{fig:fig4}.B), separating them by anatomical layer and brain region. However, 
since we recorded unequal number of cells in the different layers (see Supplementary Figure \ref{fig:S1}), we also accounted for the different numbers of cells and recordings per layer
and estimated chance-level expectations for the connectivity style counts in each layer: 
this allowed us to 
detect significant over- or under-representations of certain styles at different locations. The numbers of ``streamers'' (core), ``staff helpers'' (regular core-skin) and ``callers'' (periphery) profiles were compatible with chance levels at all the recorded locations.
We only detected over-representations (green upward triangles) of ``free-lancer helpers'' (bursty core-skin)  in Stratum Radiatum (SR) of CA1 and 
Layer II of medial Enthorinal Cortex, and an under-representation (red downward triangle) of this style
in Layer III of medial Enthorinal Cortex (see \textit{Discussion} for possible interpretations). 
These moderate deviations from chance levels suggest that the connectivity styles adopted by different neurons (and thus their centrality in the core-periphery architecture of information sharing networks) are only poorly affected by their anatomical location in the hippocampal formation circuit, in apparent contrast with the widespread belief that the ``hubness'' of neurons should be strongly determined by structural and developmental factors \cite{Cossart:2014cu}.

We found however a stronger inter-relation between cell type and connectivity styles (Figure \ref{fig:fig4}.D, bottom). We still found representatives of any of the connectivity styles among both excitatory and inhibitory neurons. However we found that the fraction of inhibitory (resp., excitatory) neurons among the core neurons was significantly above (resp., below) chance level. Conversely, the fraction of inhibitory (resp., 
excitatory) neurons among the peripheral neurons was significantly below 
(resp., above) chance level (see \textit{Discussion}).

\subsection{Connectivity profiles are network-state dependent and not only node-dependent}

We have computed connectivity profiles per neuron \textit{and} per network state, in order to enable the detection of 
a possible network-state dependency of the temporal properties of the neurons connectivity. 
However, the state-specificity of this computation does not prevent a priori a neuron to always assume the same connectivity profile across all possible network states to which it participates. It is thus an open question, whether connectivity profiles are only
node-dependent (for a given neuron, the same in every state) or, more generally, 
state-dependent (for a given neuron, possibly different across different network states). 

\begin{figure}[htb]
\centerline{\includegraphics[width=\textwidth]{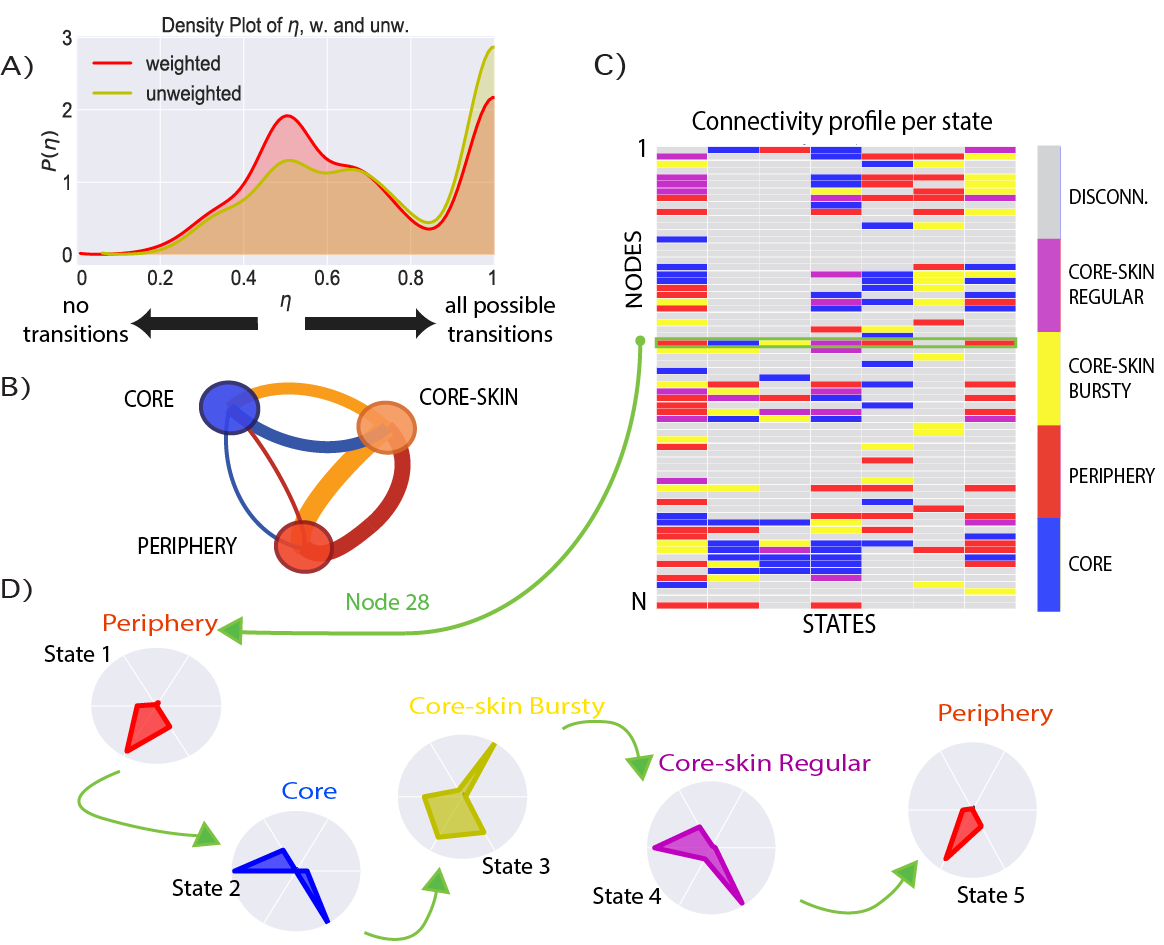}}
\caption{\textbf{Network-state specificity of connectivity profiles.}
\textbf{A)} Histogram of the connectivity-style-switching probability $\eta$
of nodes, in the weighted (red) and unweighted (yellow) cases, computed for each neuron in each recording. $\eta$ is defined 
(see \textit{Methods}) as the ratio 
between the number of transitions between different connectivity styles in subsequent states of a neuron during the whole recording 
and the maximum value between the total number of connectivity styles and number of network states observed in the recording.
\textbf{B)} Visualization as a weighted, directed graph of the cross-network-state connectivity style transition matrix. The color of an edge corresponds 
to the color of the node from which it 
emanates. The width of an edge from 
connectivity style $A$ to connectivity style $B$ 
is proportional to the probability of a neuron to switch from 
a connectivity profile of connectivity style $A$
in a state to a profile of connectivity style $B$ in the next state. 
\textbf{C)} Matrix representation of the changes
in connectivity styles of the neurons when the network changes state. Each element $(s,n)$ of the matrix is coloured according to the connectivity style to which node $n$'s connectivity profile belongs in state $s$. 
\textbf{D)} Connectivity profile transitions of 
node $n=28$, whose evolution is characterized by transitions between profiles belonging to different styles, highlighting how a given node can display radically different dynamical behaviours in different states.}
\label{fig:fig5}
\end{figure}

To answer this question, we checked whether network state transitions are associated or not to connectivity style modifications at the level of individual neurons. We found that changes in the connectivity style of a neuron upon a change of state are the norm rather than the exception. 
We computed for every neuron the index $\eta$, quantifying the diversity of connectivity styles that a neuron assumes across the different network state transitions occurring during a recording. 
Such an index is bounded in the range $0 \le \eta \le 1$ and assumes the null value if the neuron always remains in the same connectivity style (no style transitions); it takes the 
unit value if the neuron 
changes connectivity style every time that a network state transition occurs, and it 
assumes intermediate values if style transitions occur for some of the network state transitions but not for all 
(see \textit{Methods}). The distribution of the observed values of this $\eta$ index is shown in Figure \ref{fig:fig5}.A. This distribution is 
bimodal, with a first peak occurring around 
$\eta\sim 0.5$ and a second at $\eta \sim 1$ (computed according to either 
unweighted or weighed network state transitions). 
This means that almost no neuron was associated to a network state-independent, fixed connectivity style. 
On the contrary, a large number of neurons changed connectivity style in at least roughly half of the network state transitions 
(first mode of the distribution), and many changed style at almost each state transition (second mode of the distribution of $\eta$).

Figure \ref{fig:fig5}.B is a graph-representation of the transition matrix between connectivity
styles, computed over all the observed connectivity style transitions. 
We plot here a weighted, directed graph, in which  the two connectivity styles ``regular'' and ``bursty core-skin'' have been merged for simplicity into a single category. 
The  width of each colored edge corresponds to the value of the transition rate from the class of the same color. 
The transition rate is defined as the probability that a node characterized by one of the 
three connectivity styles in one network-state, switches to one of the other two connectivity styles in the successive network-state. 
Consistently with Figure \ref{fig:fig4}.C, we find that edges of large weight connect the core-skin to the periphery and to the core
in both directions: high transition rates are found between these styles. The edges connecting core and periphery are sensibly smaller.
We can thus conclude that it is largely more likely that neurons of a \emph{core} connectivity profile switch to a \emph{core-skin} profile in the following network state than to a \emph{peripheral} profile, and vice versa:
 direct transitions between core profiles and periphery profiles are not very likely to occur, although they are not impossible. More complete transition graphs and tables (including as well the core-skin class separation into ``bursty'' and ``regular'', and the ``junk'' classes) are shown in Figure \ref{fig:S5}.
Individual neurons can thus float through the core-periphery architecture, descending from core towards periphery and  
ascending back into the core, via crossing the core-skin styles.

The overall behavior of neurons switching styles between states is illustrated in Figures \ref{fig:fig5}.C and \ref{fig:fig5}.D for a specific recording. 
The former is a matrix whose element $(i,s)$ has the color of the connectivity style exhibited by node $i$ in state $s$. 
In this plot we highlight the row corresponding to the state-wise evolution of a specific selected node, 
whose successive connectivity profiles in the successive network state ares shown in Figure \ref{fig:fig5}.D.

\subsection{Connectivity profiles only poorly depend on firing rate}

In our recordings we observed a diversity of firing rates between different single units. While the median firing rate was close to $\sim 1$ Hz, some neurons had, at specific moments, high rates approaching $40$ Hz, despite the anesthesia conditions. Importantly, firing rates were also changing across the different network states. To check whether variations of temporal connectivity features compiled in the state-specific connectivity profiles of different neurons could simply
be explained by these firing rate variations, we constructed scatterplots of state-specific coreness, liquidity, strength, number of activations, Fano factors and total time of activation for the different neurons against their firing rate, averaged over the corresponding state. 
These scatter plots are shown in Supplementary
Figure \ref{fig:S4}.C. 

The degree of correlation between firing rate and connectivity features was at best mild. Cells with larger than median firing rate tended to have slightly less liquid neighborhoods on average, but the range of variation was broad and largely overlapping with the one of liquidity values for lower firing rate neurons. Other features, such as total time of activation, displayed an even weaker dependence on firing rate, while some others,
such as the connectivity number, tended to be inversely correlated with firing, but, once again, in a rather weak manner (the correlation coefficient is $\rho=-0.14$).

Overall, the analyses of Figure \ref{fig:S4}.C indicate that the structure of information sharing networks is shaped by the coordination of firing between units more than by the firing rates of individual cells. 

\subsection{Relations between connectivity styles and active information storage}

We focused in this study on the dynamics of links of information sharing. Information sharing can be seen as a generalized measure of cross-correlation between the firing activity of two different single units (see \emph{Methods}). In \cite{Clawson2019}, we analyzed as well \emph{active information storage} \cite{Lizier:2012db}, a complementary measure, which can be seen as a generalized auto-correlation of the firing activity of a single node (or, equivalently, to a self-loop of information sharing connectivity). While information sharing quantifies the amount of information in common between the present activity of a neuron and the past activity of another, information storage is meant to capture the amount of information in the present activity of a neuron which was already conveyed by its past activity (see \emph{Methods}). In this sense, neurons with a large value of information storage would functionally act as ``memory buffers'' repeating in time the same information content (cf. \emph{Discussion}). In \cite{Clawson2019} we found that information storage, as information sharing, is state-dependent and that some cells, that we called therein ``storage hubs'' display particularly large values of storage. We checked here whether high storage cells manifest some preferential connectivity style. 

We show in Figure \ref{fig:S6}, distributions of information storage values, separated according to the connectivity style assumed by different neurons in each different network state. We found that high storage is not associated exclusively to specific styles of connectivity. On the contrary, cells with high storage could be found for any of the connectivity styles. The overall larger storage values where found for core streamers cells.
However, we could find high storage cells even among the core-skin helpers and the periphery callers. Interestingly, distributions of storage for core-skin and periphery styles were bimodal, including a peak at high-storage values (see \emph{Discussion} for possible interpretations).


\section{Discussion}\label{discussion}

We have here described the internal organization of assemblies of neurons dynamically exchanging information through time, using a temporal
network framework and developing adequate tools
for their analysis. 
In line with our previous study \cite{Clawson2019}, we found a coexistence of organization, such as the existence of discrete network states, and freedom, such as the liquid continuous reorganization of network neighborhoods within each state. The rich information sharing dynamics we revealed in our anesthesia recordings in hippocampus and enthorinal cortex cannot be explained merely in terms of transitions between global brain states (here, alternations between ``REM-sleep''-like THE epochs and SO epochs). On the contrary, we found that each of the global states gives access to wide 
repertoires of possible networks of information sharing between neurons. These network states repertoires are largely global state-specific and are robustly identified using both weighted and unweighted network characterizations.

We extracted here functional connectivity networks evaluating information sharing ---i.e., time-lagged mutual information--- between the firing of pairs of single units. This choice was motivated by the fact that mutual information is relatively simpler to evaluate than other more sophisticated and explicitly directed measures of information transfer \cite{Schreiber2000}, which require larger amounts of data to be properly estimated. Using a simpler measure was thus better compliant with our need to estimate connectivity within short windows, to give rise to a temporal network description. Although mutual information is a symmetric measure, the introduction of a time-lag makes our metric ``pseudo-directed'' because information cannot causally propagate from the future to the past. Therefore, sharing of information between the past activity of a node $i$ and the present activity of another node $j$ could be indicative of a flow of information from $i$ to $j$. When computing time-lagged mutual information between all directed pairs of single units in our recording, we found however only a very mild degree of asymmetry. As shown in Figure \ref{fig:S5}, most strong information sharing connections were very close to symmetric, indicative of time-lagged mutual information peaking near zero-lag for these connections. This finding corresponds to the early intuition that members of a cell assembly share information via their tightly synchronized firing \cite{Abeles1982, Hebb1949}. Figure \ref{fig:S5} also shows that some of the more weakly connected pairs of neurons displayed a varying degree of directional asymmetry, with the strongest unbalance found for cells with a core connectivity style. However, the strengths of nodes with unbalanced pseudo-directed sharing were orders of magnitude smaller than for balanced nodes. Our choice to ignore asymmetries appears thus well justified, as well as the choice to use lagged mutual information rather than more complex directed metrics.

A striking architectural property of the measured information sharing networks was their 
core-periphery architecture. Such architecture was preserved at every time-frame (as shown in Figure \ref{fig:fig2}.B), although the coreness of individual neurons changed smoothly through time even within states. This preservation of a global functional architecture despite the variation of the specific realizations and participant neurons is reminiscent of ``functional homeostasis'' observed in highly heterogeneous circuits that must nevertheless perform in a stable and efficient manner a crucial function \cite{Marder:2006gt, Kispersky2011}. What is important is that, at any time, a functional separation between ``streamers'', ``helpers'' or ``callers'' exist, however which neurons specifically assume these roles and the exact location of these neurons within the anatomical circuit (cf. Figure \ref{fig:fig4}.D) appear to be less important.

Core-periphery organizations have been found in many social, infrastructure, 
communication and information processing networks, with various types of coreness profiles, i.e.,
strong or gradual separations between the
innermost core and the most peripheral nodes
\cite{Rossa,Kojaku:2018,Rombach2014,Holme:2005}.
The dynamics of this type of structures has however
barely been tackled in temporal networks
\cite{galimberti2018mining}.
In neuroscience, such architecture has been 
identified in larger-scale networks of inter-regional functional connectivity, during both task and rest \cite{bassett2013task, avena2017, pasquale2017, battiston2018}, in line with the general idea that cognition requires the formation of integrated coalitions of regions, merging information streams first processed by segregated sub-systems \cite{Tononi:1998vk, Deco:2015ge}. Here, we find that a similar architecture is prominent at the completely different scale of networks between single units within local micro-circuits in the hippocampal formation. The core-periphery architecture of information sharing thus reflects dynamic information integration and segregation. Cells belonging to the core form an integrated ensemble, within which strong flows of information are continually streamed and echoed. At the opposite end of the hierarchy of connectivity styles, peripheral neurons are segregated and perform transient ``calls'' toward ``streamer'' neurons in the core to get updated and share their specific information contents.

Following \cite{Buzsaki2010}, a ``syntax of information processing'' would be enforced by the spatio-temporal organization of neuronal firing. Our dynamic core-periphery architecture is also ultimately determined by the coordinated firing patterns of many neurons. The fact that these coordinated firing patterns invariantly translate into networks with a core-periphery architecture can be seen as a proxy for the existence of a syntactic organization of information flows. In this self-organized syntax, peripheral neurons are in the ideal functional position to act as ``readers'' of the contents streamed by the integrated cell assembly formed by core neurons. Transitions between network states may act as the equivalent of bar lines, parsing the flow of information into ``words'' or longer sentence blocks, spelled with an irregular tempo. In the vision of \cite{Buzsaki2010}, roles such as the one of reader or chunking into informational words would be a by-product of neuronal firing organization. We stress thus here once again that even in our case the real ``stuff'' out of which sharing functional networks are made are spiking patterns at the level of the assembly.  The network representation provides however a natural and intuitive visualization of the dynamics of these patterns. Coordinated firing events ---and not single neuron firing properties, cf. the poor correlations in Figure \ref{fig:S4}--- are mapped into network structures and changes of firing pattern into changes of connectivity style within the network. The logical syntax of assembly firing is thus translated into a dynamic topological architecture of sharing networks, which is easier to study and characterize.

In order to interpret the functional role of this emergent core-periphery organization, one should elucidate the information processing roles played by individual neurons within this network organization. An entire spectrum of possible roles may exist, mirroring the smooth separation subsisting between core and periphery. Communication between core and periphery can be helped by core-skin neurons, which sporadically connect to the core, transiently expanding its size. Speculatively, such breathing of the integrated core may be linked to fluctuating needs in terms of information processing bandwidth, analogous to dynamic memory allocation in artificial computing systems. The possible algorithmic role of neurons with different connectivity styles could be investigated using metrics from the partial information decomposition framework \cite{Wibral:2017im}. Beyond pseudo-directed sharing, it may be possible to identify: neurons acting as active memory buffers displaying large information storage \cite{Lizier:2012db}; neurons actively transfering information to others, associated to large transfer entropy \cite{Schreiber2000}; or, yet, neurons combining received inputs into new original representations, associated to positive information modification values \cite{Lizier:2013hu}. As previously said, we could at least quantify information storage.  In particular, as shown in Figure \ref{fig:S6}, sub-groups of core-skin helpers and periphery callers existed with high storage values, approaching the ones of storage hubs within the core. Therefore, at least some of the core-skin or periphery cells transiently connected to the core may play the role of ancillary memory units, ``flash drives'' sporadically plugged-in when needed to write or read specific information snippets. Unfortunately, the number of coordinated firing events observed in our recordings was not sufficient for a reasonable estimation of transfer or modification, beyond storage. We can nevertheless hypothesize that 
core neurons are the work-horses of information modification, shaping novel informational constructs via their integrated firing.

We found that the bursty core-skin style ---the ``freelancer helper'' role--- was over-represented in SR of CA1 and Layer 2 of EC. SR receives inputs from hippocampus CA3 which are believed to be linked to retrieval of previously encoded associations \cite{hasselmo2013, schomburg2014}. CA3, an associative memory module crucial for retrieval, receives inputs on its turn from layer 2 of EC. The over-representation of the bursty core-skin style in SR and layer 3 of EC  may thus be compliant with our intuition of ``helpers'' as additional on-demand storage resources, plugged to the core when needed to read the specific contents buffered by their spiking. On the contrary, layer 3 of EC, which sends inputs to SLM of CA1, would rather mediate encoding than retrieval \cite{douchamps2013, schomburg2014}. Correspondingly, in layer 3 of EC the over-representation of freelancer helpers observed in layer 2 is replaced by an under-representation, consistently with the complementary functions that inputs from these EC layers are postulated to play.

We also found an excess of inhibitory interneurons among core style cells (Figure \ref{fig:fig4}.D). Inteneurons tended to exhibit larger firing rates than excitatory cells, but we showed that coreness is not significantly affected by firing rate (Figure \ref{fig:S4}.C). Therefore, this excess is rather to be linked to a key functional role of interneurons in mediating cell assembly formation. Their central position within the core of the information sharing network may allow them to efficiently control the recruitment of new neurons into the integrated core and orchestrate their coordinated firing, as already discussed in the literature \cite{Bonifazi:2009in,quilichini2012}.

Remarkably, however, in a majority of cases, the connectivity style adopted by a neuron was only poorly affected by its anatomical localization or by its cell type. Apart from the few exceptions just discussed above, the distribution of the different styles through the different anatomical regions and layers was indeed close to chance levels. In particular, the fact of being localized within a specific layer of CA1 or EC did not affect in a significant way the probability of belonging to the core or to the periphery. Furthermore, a majority of neurons switched between different styles when network state changed (Figure \ref{fig:fig5}), indicating that connectivity styles and, in particular, core membership are not hardwired. In contrast, it is often thought that the function that a neuron plays is affected by its individual firing and morphology properties, as well as by details of its synaptic connections within the circuit \cite{booker2018}. In this dominating view, functional hubness would thus be the garland reserved to a few elite cells, selected because of their extreme technical specialization, largely determined by their developmental lineage \cite{Cossart:2014cu}. Here ---further elaborating on \cite{Clawson2019}--- we propose a more ``democratic'' view in which core membership and, more in general, connectivity style would be dynamically appointed, such that the total number of neurons that are elected into the core at least once is much larger than the currently active core members at specific times. The core composition can indeed be radically reorganized when the network state switch and can fluctuate between alternative majorities of hippocampal or enthorinal cortex neurons (Figure \ref{fig:fig2}.D). 

Such a democratic system implies a primacy of collective dynamics at the neuronal population level, flexibly shaping coordinated firing ensembles, on technical specialization and ``blood'' origins at the single neuron level. Switching network states may reflect transitions between alternative attractors of firing dynamics \cite{Amit:1989:MBF:77051} implementing alternative firing correlations. Discrete state switching coexist however with more liquid fluctations of core attachment, which would rather suggest a complex but not random dynamics at the edge of instability \cite{Maass:2002kf, Marre:2009ho}. State transitions and flexible core-periphery reorganization are also poorly determined by global oscillations, despite the important role played by oscillations in information routing \cite{Buzsaki2006, Womelsdorf:2007hj}. Even if global oscillatory modes do not ``freeze'' information sharing patterns they nevertheless affect them, with different oscillatory states giving rise to alternative repertoires of possible information sharing networks (Figure \ref{fig:fig3}.D). Computational modelling of spiking neural circuits may help in the future
to reach a mechanistic understanding of
how ongoing collective oscillations interact with discrete attractor switching, metastable transients and plasticity to give rise to liquid core-periphery architectures of information sharing.

We have started here applying a temporal network language to describe the internal life of cell assemblies along their emergence, expansion and contraction and sudden transformations. For methodological convenience --- possibility to use long time windows for network estimation--- we focused however on anesthesia, which is a condition in which intrinsic information processing is not suppressed but less functionally relevant than during behavior. Our method could however be extended in perspective to recordings in pathological conditions --- e.g., epilepsy, in which intrinsic assembly dynamics is altered \cite{FeldtMuldoon:2013iq}---, relating temporal network properties alterations to the degree of cognitive deficits  or even, in perspective, during actual tasks. To cope with the much faster behavioral time-scales, information sharing networks could be estimated first in a state-resolved manner, by pooling together firing events based on the similarity of the conditions in which they occur --- e.g., transient phase relations and synchrony levels \cite{Palmigiano:2017he} or co-activation patterns \cite{Miller:2014eq, Malvache:2016dd}--- rather than strict temporal contiguity. State-specific network frames could then be reallocated to specific times, depending on which ``state'' the system is visiting at different times, reconstructing thus an effective temporal network with the same time-resolution as the original recordings (see e.g. \cite{Thompson:2017} for an analogous approach used at the macro-scale of fMRI signals). In this way it would become possible to link temporal network reconfiguration events to actual behavior, probing hence their direct functional relevance.



\section{Methods:}\label{methods}

\subsection{Animal surgery}

We used in this work a portion of the data (12 of 18 experiments) initially published by \cite{Quilichini:2010bt}, which includes local field potentials (LFPs) and single-unit recordings obtained from the dorsomedial Enthorinal Cortex of anesthetized rats. Seven additional simultaneous recordings in both mEC and dorsal HPC under anesthesia were included in this study, previously analyzed by \cite{Clawson2019}. Details of anatomical locations and numbers of cells included can be found in Supplementary Figure \ref{fig:S1}.

All experiments were in accordance with experimental guidelines approved by the Rutgers University (Institutional Animal Care and Use Committee) and Aix-Marseille University Animal Care and Use Committee. We performed experiments on 12 maleSprague-Dawley rats (250 to 400 g; Hilltop Laboratory Animals) and 7 male Wistar Han IGS rats (250 to 400 g; Charles Rivers Laboratories). We performed acute experiments anesthetizing these rats with urethane (1.5 g/kg, intraperitoneally) and ketamine/xylazine (20 and 2 mg/kg, intramuscularly), with additional doses of ketamine/xylazine (2 and0.2 mg/kg) being supplemented during the electrophysiological recordings to avoid recovery from anesthesia. The body temperature was monitored and kept constant witha heating pad. The head was secured in a stereotaxic frame (Kopf)
and the skull was exposed and cleaned. Two miniature stainless steel screws, driven into the skull, served as ground and reference electrodes. To reach the mEC, we performed one craniotomy from bregma: $-7.0$ mm anteroposterior (AP) and $+4.0$ mm mediolateral (ML); to reach the CA1 area of HPC, we performed one craniotomy from bregma: $-3.0$ mm AP and $+2.5$ mm ML. 
The probes were mounted on a stereotaxic arm. We recorded the dorsomedial portion of the mEC activity using a NeuroNexus CM32-4x8-5 mm-Buzsaki32-200-177 probe (in eight experiments), a 10-mm-long Acreo single-shank silicon probe with 32 sites (50mm spacing) arranged linearly (in five experiments), or a NeuroNexus H32-10mm-50-177 probe (in five experiments), which was lowered in the EC at 5.0 to 5.2 mm from the brain surface with a 20 degrees angle. We recorded HPC CA1 activity using a H32-4x8-5mm-50-200-177 probe (NeuroNexus Technologies) lowered at 2.5 mm from the brain surface with a 20 degree angle (in four experiments), a NeuroNexus H16-6mm-50-177 probe lowered at 2.5 mm from the brain surface with a 20 degree angle (in two experiments).
The on-line positioning of the probes was assisted by the presence of unit activity in cell body layers and the reversal of theta oscillations when passing from L2 to L1 for the mEC probe and the presence in SP of either unit activity or ripples (80 to 150 Hz) for the HPC probe.

At the end of the recording, the animals were injected with a lethal dose of pentobarbital Na (150 mg/kg, intraperitoneally) and perfused intracardially with 4\% paraformaldehyde solution. We confirmed the position of the electrodes (DiI was applied on the back of the probe before insertion) histologically on Nissl-stained 40-mm section.

\subsection{Data collection and spike sorting}

Extracellular signal recorded from the silicon probes was amplified ($1000x$),
bandpass-filtered (1 Hz to 5 kHz), and acquired continuously at 20 kHz with a 64-channel DataMax System (RC Electronics ora 258-channel Amplipex) or at 32 kHz with a 64-channel DigitalLynx (NeuraLynx at 16-bit resolution). We preprocessed raw data using a custom-developed suite of programs \cite{csicsvari1999}. After recording, the signalswere downsampled to 1250 Hz for the LFP analysis. Spike sorting wasperformed automatically using KLUSTAKWIK (http://klustakwik.sourceforge.net \cite{harris2000}]), followed by manual adjustment of the clusters, with the help of autocorrelogram, cross-correlogram (CCG), and spike waveform similarity matrix (KLUSTERS software package;\verb+https://klusta.readthedocs.io/en/latest/+ \cite{hazan2006}) After spike sorting, we plotted the spike features of units as a function of time and discarded the units with signs of significant drift over the period of recording. Moreover, we included in the analyses only units with clear refractory periods and well-defined clusters. 

Recording sessions were divided into brain states of THE and SO periods. The epochs of stable theta (THE) or slow oscillations (SO) were visually selected from the ratios of the whitened power in the THE band ([3 6] Hz in anesthesia) and the power of the neighboring bands of EC layer 3 LFP, which was a layer present in all the 18 anesthesia recordings. 

We determined the layer assignment of the neurons from the approximate location of their somata relative to the recording sites (with the largest amplitude unit corresponding to the putative locationof the soma), the known distances between the recording sites, and the histological reconstruction of the recording electrode tracks. To assess the excitatory or inhibitory nature of each of the units, we calculated pairwise crosscorrelograms between spike trains of these cells. We determined the statistical significance of putative inhibition or excitation (trough or peak in the $[+2:5]$ ms interval, respectively) using well-established nonparametric test and criteria used for identifying monosynaptic excitations or inhibitions \cite{Quilichini:2010bt}, in which each spike of each neuron was jittered randomly and independently on a uniform interval of $[-5:5]$ ms a thousand times to form 1000 surrogate datasets and from which the global maximum and minimum bands at 99\% acceptance levels were constructed.

\subsection{Information sharing networks}

We estimated time-resolved information sharing networks following the same procedures as \cite{Clawson2019}.  In particular, we used \emph{time lagged Mutual Information} between spike trains as a measure of the amount of \emph{shared information}. Information sharing between each pair of neurons $(i,j)_{i,j=1,\dots,N}$ is defined  as the time lagged Mutual Information $ MI[i(t),j(t-\lambda)]$ between the firing patterns extracted for the relative neurons in a $10$ seconds long time-window:
\begin{equation}
    I_{shared}(j\xrightarrow{}i)=\sum_\lambda MI[i(t),j(t-\lambda)]
\end{equation}{}
\begin{equation}
    I_{shared}(i\xrightarrow{}j)=\sum_\lambda MI[j(t),i(t-\lambda)]
\end{equation}{}
where the time-lag $\lambda$ varies in the range $0\leq\lambda\leq   0.5 T^\theta$, $T^\theta$ being the phase of the THE cycle (THE oscillations have higher frequency than slow oscillations). The information sharing network is defined by an adjacency matrix $A(i,j)_{i,j=1,\dots,N}$, whose element $A(i,j)$ corresponds to $I_{shared}(i\xrightarrow{}j)$. 
It is therefore a weighted, directed network. The time-window is then shifted by $1$ second to extract the information sharing network at the next time-step, assuring a $90\%$ overlap with the previous time-window for the computation of the information theoretic measures. 

In order to study the effect of asymmetries in the adjacency matrices of our information sharing networks we compute the \emph{asymmetry coefficient} $\Delta S\%$ for each neuron in each state $h$. 
defined as follows:
\begin{equation}
    \Delta S\% ^i_h=\frac{s^{i,h}_{out}-s^{i,h}_{in}}
    {s^{i,h}_{out}+s^{i,h}_{in}},\quad \forall i=1,\dots,N  \ .
\end{equation}
 
 $\Delta S\%^i_h$ thus represents the ratio between the amount of information that neuron $i$ outputs to other neurons during state $h$ ($s^{i,h}_{out}=\sum_{t=1,\dots,T^h}\sum_j w_{ij}$) and the amount of information that is conveyed to neuron $i$ by its neighbors during state $h$ ($s^{i,h}_{in} = \sum_{t=1,\dots,T^h}\sum_j w_{ji}$). ${\Delta S\%}^i_h=1$ means that for neuron $i$ in state $h$, $s^{i,h}_{in}=0$, therefore it is a perfect \emph{sender} of information, whereas if ${\Delta S\%}^i_h=-1$,  $s^{i,h}_{out}=0$ and the neuron is therefore a perfect \emph{receiver}. If ${\Delta S\%}^i_h=0$ then $s^{i,h}_{out}=s^{i,h}_{in}$ and therefore the amount of information that neuron $i$ conveys to its neighbors during the state $h$ is equal to the amount of information that it receives. As shown in Supplementary Figure \ref{fig:S7}, the asymmetries in the adjacency matrices of the information sharing networks have a negligible effect. We thus neglect the aforementioned asymmetries and consider the networks to be undirected.

\subsection{Active Information Storage}

Information sharing can be seen as a generalized form of cross-correlation where the time-lagged mutual information capture all types of linear and nonlinear correlations. Analogously, one can evaluate a generalized auto-correlation functional, known under the name of active information storage \cite{Lizier:2012db}. Following \cite{Clawson2019}, we evaluated active information storage of a given neuron, within a given time-window as:
\begin{equation}
    I_{storage}(i)=\sum_\lambda MI[i(t),i(t-\lambda)]
\end{equation}{}
where the time-lag $\lambda$ varies once again in the range $0\leq\lambda\leq   0.5 T^\theta$. The values of storage whose distributions are shown in Figure \ref{fig:S6} are averages for each neuron over all the time-windows whose associated connectivity profile much a specific connectivity style (see later for definitions of connectivity profiles and styles).

\subsection{Network liquidity}
To quantify
how much, or how little, the connections of a the node's neighborhood change from a time-step to the next, 
we compute the cosine similarity (for weighted networks) and the Jaccard index (for unweighted ones)
between the successive neighborhoods of the node.
Both measures have been intensively used to study the stability or instability of connections in temporal networks \cite{Darst,STEHLE2013604,gelardi:2019}. The cosine similarity  between a node $i$'s neighborhoods at times $t-1$ and $t$ is defined as follows:
\begin{equation}
    \Theta^i(t)\equiv\frac{\sum_{j} w_{ij}(t-1)w_{ij}(t) }{\sqrt{\sum_{j} w_{ij}(t-1)^2}\sqrt{\sum_{j} w_{ij}(t)^2}} 
\end{equation}{}
where 
 $w_{ij}(t)$ is the weight of the link 
 between nodes $i$ and $j$ at time $t$.

The local \emph{Jaccard index} for node $i$ at time $t$ (Figure \ref{fig:fig1}.C, on the left) is defined as: 
\begin{equation}
    J^i(t)\equiv\frac{|\nu^i(t-1)\cap \nu^i(t)|}{|\nu^i(t-1)\cup \nu^i(t)|}\quad,\quad J^i(t)\in[0,1]
\end{equation}{}
where $\nu^i(t-1)$ and $\nu^i(t)$ are the neighborhoods of node $i$ at times $t-1$ and $t$, respectively.
$J^i(t)=0$ when $\nu^i(t-1)$ and $\nu^i(t)$ are two disjoint sets of nodes, while $J^i(t)=1$ when the two sets are identical. 

Both the cosine similarity and the Jaccard index are thus indicators, for weighted and unweighted networks respectively, 
of a node's tendency to disrupt and create new edges from a time-step to the next ($J^i(t)\sim 0$ implies high liquidity), or to rather maintain stable connections within subsequent time-steps ($J^i(t)\sim 1$ implies low liquidity). 

At each time $t$ the outcome of the computation are two vectors of dimension $N$: 
$\mathbf{\Theta}(t) = (\Theta^1(t), \ldots, \Theta^N(t))$ and 
$\mathbf{J}(t) = (J^1(t), \ldots, J^N(t))$.

\subsection{Coreness} 

We consider the definition of \emph{coreness} $C^i$ of node $i$ in a network introduced by \cite{Rossa}. To compute the coreness of neurons at each time frame we used the implementation of the method available at
\texttt{https://core-periphery-detection-in-networks.readthedocs.io/en/latest/index.html}.
When computed for a static, unweighted, undirected network, the coreness $C^i$ of node $i\in[1,N]$ is a real number between $0$ and $1$, 
 interpreted as follows: when $C^i\sim 1$ the node belongs to the \emph{core} of the network; 
 when $C^i\simeq 0$ the node belongs to the network's \emph{periphery}; if $C^i=0$ the node is disconnected from the network, i.e., 
 is not linked to any other node. 
 
 Analogously to what was done for the local cosine similarity and the local Jaccard index, at each time step $t\in[0,T]$ we compute the coreness for each node $i\in[1,N]$ for both the unweighted
 and weighted networks: 
 at each time step we therefore obtain two $N$-dimensional feature vectors, one for the weighted network 
 $\{C_w^i(t);\ i = 1,\cdots,N \}$
 and one for the unweighted one
 $\{C^i(t);\ i = 1,\cdots,N \}$. 

\subsection{Network states} 

The two pairs of feature vectors computed at each time $t$, two vectors for the unweighted case and two for the weighted one, 
are merged into two $2\times N$-dimensional vectors, one containing the unweighted features (Jaccard and unweighted coreness) and 
the other formed by the weighted ones (cosine similarity and weighted coreness). 
For each recording we have therefore two series of length $T$ ($T$ varies for different recordings), 
of weighted and unweighted feature vectors. In order to investigate the existence of recurring epochs of stable (correlated) liquidity and topology configuration of the network (the \emph{network states}) we perform a $K$-means clustering on the feature vector time series. 
As a result of the clustering we obtain for each recording two sets (one for the weighted analysis, one for the unweighted one)
of network states of specific liquidity (Jaccard index, for unweighted networks, or cosine similarity, for weighted networks) and 
core-periphery organization (unweighted or unweighted coreness) of the system (Figure \ref{fig:fig3}.A).

\subsection{Clustering connectivity profiles: connectivity styles}  

As shown in Figure \ref{fig:fig4}.B, we 
represent the connectivity behaviour of a neuron in a network state with a radar plot whose radial axis correspond to the six features defined in section 
Results, which are averages of temporal properties
over the network state. 
We therefore compute one six-dimensional \emph{connectivity profile} for each neuron, in each network state, for each recording. 
We investigate the existence of typical connectivity profiles across different recordings: this
corresponds to searching for common radar plot 
shapes for neurons of different network states of different recordings. 
We therefore perform a K-means clustering on all 
the connectivity profiles computed for each neuron, in each network state, in each recording: we obtain  four cluster of connectivity profiles, that we name \emph{connectivity styles}. These connectivity styles
are shown in Figure \ref{fig:fig4}.B and interpreted in the Results
and Discussion sections.

\subsection{Unsupervised classification for soft labeling of connectivity profile}

We use the K-nearest-neighbor unsupervised classification method, trained on the connectivity style labels of all the connectivity profiles, to obtain a soft classification of each neuron's connectivity profile. Each neuron's state-averaged connectivity profile's soft label is represented by a $4-$dimensional normalized vector: 
each component of this vector represents 
the probability of the connectivity profile to belong to the corresponding connectivity style 
(core, periphery, bursty core-skin, regular core-skin).
The connectivity styles shown in Figure \ref{fig:fig4}.B correspond to the connectivity profiles of maximum probability to belong to the connectivity style of the corresponding colour, i.e. of maximum soft label component of the corresponding connectivity style. The soft classification of connectivity profiles is graphically represented in Figure \ref{fig:fig4}.C by plotting the connectivity profiles' vectors of connectivity style soft labels in a $3-D$ space. 
In order to achieve a clearer visual representation, the probabilities of belonging to the regular core-skin or the bursty core-skin connectivity styles have been summed into a single component of the soft label vector.

\subsection{Cross-network-state connectivity profile transition rate}
The cross-network-state connectivity profile transition rate $\eta^i$ of neuron $i$ 
is defined as:
\begin{equation}
    \eta^i\equiv\frac{\# ~ profile ~ transitions ~ of ~ node ~ i}{max(\# ~  global ~ classes, ~ \# ~ network-states)} .
\end{equation}{}
It quantifies the tendency of neuron $i$ to switch, often or rarely, between connectivity profiles belonging to different connectivity styles in different states: when $\eta^i=1$, neuron $i$ switches connectivity style at every change of
network state; when $\eta^i = 0$, neuron $i$ is always in the same connectivity style.

\section*{Acknowledgments}
This project has received funding from the European Union’s Horizon 2020 research and innovation programme under the Marie Sk\l odowska-Curie grant agreement No. 713750. Also, it has been carried out with the financial support of the Regional Council of Provence- Alpes-C\^ote d’Azur and with the financial support of the A*MIDEX (ANR-11-IDEX-0001-02), funded by the Investissements d'Avenir project funded by the French Government, managed by the French National Research Agency (ANR). CB and WC were supported by the M-GATE project (European Union’s Horizon 2020 research and innovation program under the Marie Sk\l odowska-Curie grant agreement no. 765549). PQ acknowledges support from FRM, FFRE, and CURE Epilepsy Taking Flight Award. DB and AB acknowledge support by the CNRS "Mission pour l'Interdisciplinarit\'e" INFINITI program (BrainTime).

\section*{Author contributions} 
Project was formulated by DB and AB, based on data and experiments by PQ and CB and previous information theoretical analyses by WC. NP, AB and DB formalised the study quantitative approaches and NP implemented and ran all the novel temporal network analyses. All authors wrote the manuscript.
\medskip


\bibliography{Arxiv-Pedreschi}

\clearpage
\newpage

\setcounter{figure}{0}
\renewcommand{\thefigure}{S\arabic{figure}}

\section{Supporting Information}

\begin{figure}[htb]
\centerline{\includegraphics[width=\textwidth]{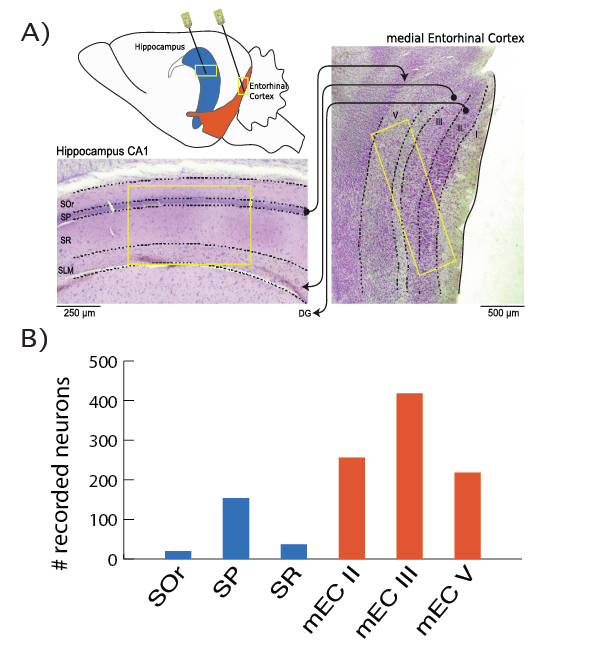}}
\caption{\textbf{A)} Simultaneous mEC/CA1 recording setup. \textbf{B)} Number of neurons recorded for each layer of each region: a majority of recorded neurons were located 
in the medial Entorhinal Cortex layers.}
\label{fig:S1}
\end{figure}

\begin{figure}[htb]
\centerline{\includegraphics[width=\textwidth]{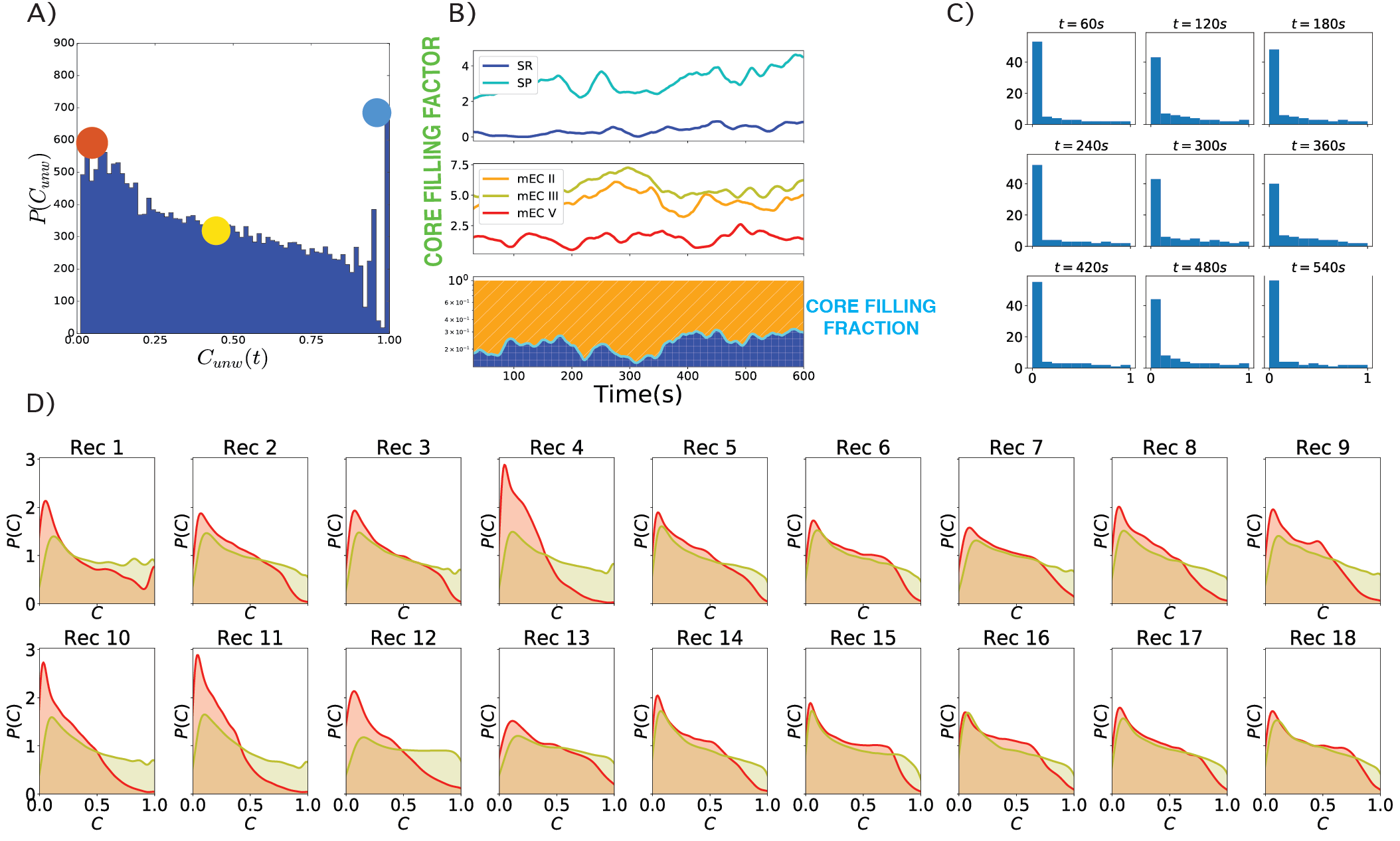}}
\caption{\textbf{A)} Histogram of the instantaneous unweighted coreness of all neurons in all time-steps for 
the same recording as in Figure \ref{fig:fig2} (unweighted analogous of Figure \ref{fig:fig2}.A).  \textbf{B)} Unweighted core-filling factors and core-filling fractions of the same recording as in Figure \ref{fig:fig2}. 
\textbf{C)} Histograms of instantaneous unweighted coreness values for $9$ different time-steps of the network's evolution.  
\textbf{D)} Density plots of the values of instantaneous weighted (red) and unweighted (yellow) coreness of all neurons at all times, separately 
for each recording.}
\label{fig:S2}
\end{figure}

\begin{figure}[htb]
\centerline{\includegraphics[width=\textwidth]{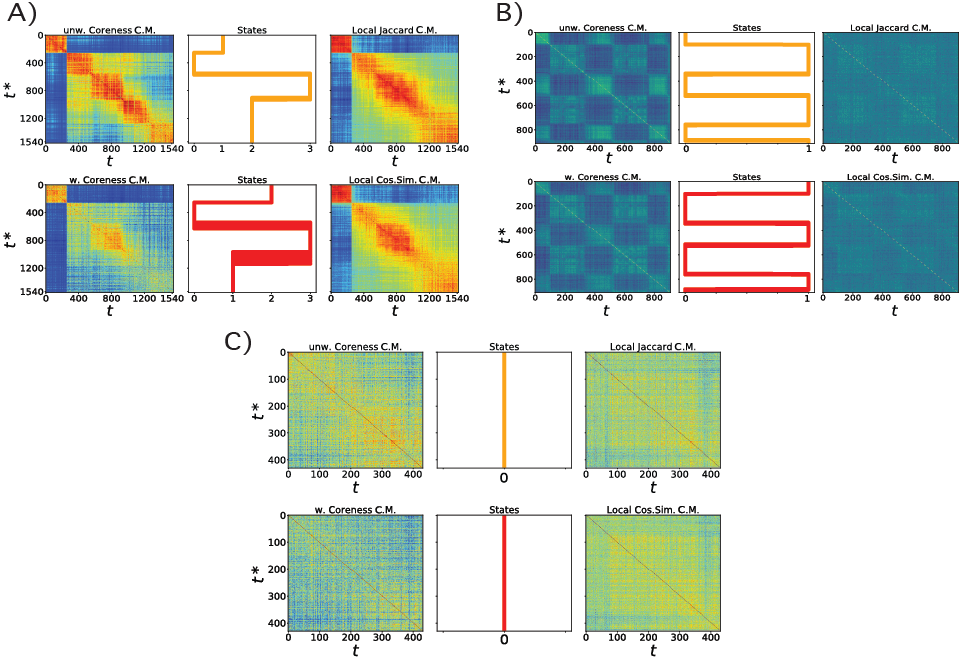}}
\caption{Recordings with different types of  
feature vectors correlation matrices and network states spectra. In each case, we show
as in Figure \ref{fig:fig3}.A. the temporal sequence
of network states extracted by the unsupervised clustering of feature vectors, 
with on both sides the correlation matrices.
Top plots correspond to unweighted features, 
bottom plots to weighted features.
\textbf{A)} Case with diagonal blocks in the
correlation matrices, with no off-diagonal blocks.
The sequence of network states is in agreement
with this structure, i.e. the network visits each network state only once. 
\textbf{B)} Case with chess-board-like correlation 
matrices. As seen from the network-state-spectra 
this recording is indeed in periodic oscillation between two states. 
\textbf{C)} Case in which no state can be 
clearly identified: this recording can be interpreted as in an extremely liquid single state, where both the core-periphery organization of the network and the neighborhoods of neurons change continuously
with no clear temporal structure.}
\label{fig:S3}
\end{figure}

\begin{figure}[htb]
\centerline{\includegraphics[width=\textwidth]{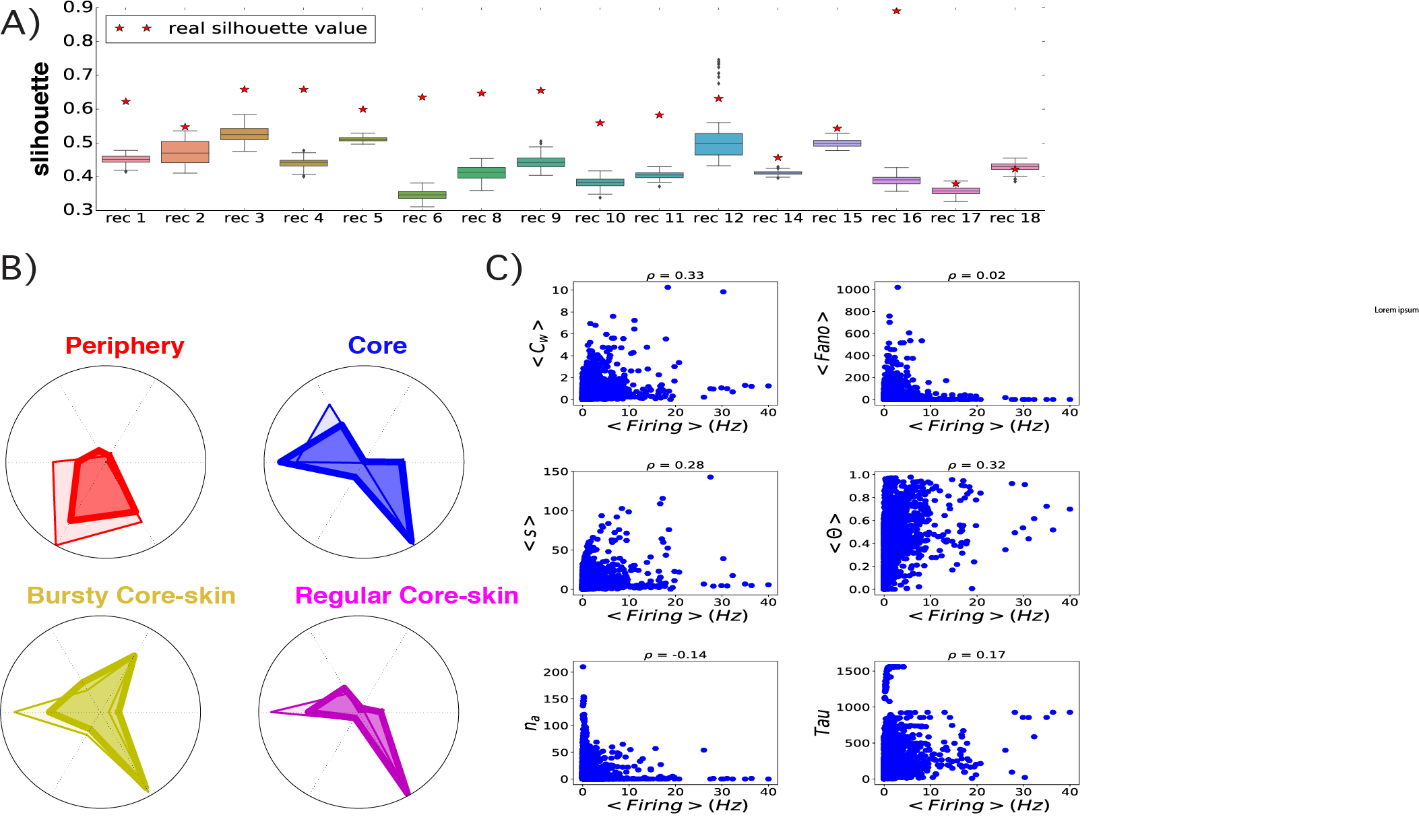}}
\caption{
\textbf{A)} Silhouette plot: the
red stars give for each recording the silhouette value for the Kmeans clustering performed on the connectivity profiles 
to retrieve the connectivity styles. 
These values are compared to distributions (boxplots) obtained for each recording by the following null model. We first reshuffle randomly the network-state labels of the time-frames of the whole recording while conserving the total length of each network-state.
We then compute the connectivity profiles on the randomized states, cluster them in order to retrieve the connectivity styles, and compute the new silhouette.
The boxplots correspond to the distribution of
these null model silhouette values, obtained
for $200$ realizations of the reshuffling.
The real silhouette values are well above the randomized distributions, 
suggesting that the definition of the discrete global states in the evolution of the information sharing liquidity and core-periphery organization 
is crucial for the analysis on the connectivity profiles and styles. 
\textbf{B)} Maximal connectivity profile of each connectivity style (lower opacity),
as shown in Figure \ref{fig:fig4}.C and  
connectivity profile of the centroid of the Kmeans clustering result (higher opacity).
We stress that the centroid of the connectivity style does not correspond to the connectivity profile of any specific a neuron, but represents the average connectivity profile of the corresponding connectivity style. 
\textbf{C)} Scatterplots between each one of the $6$ features computed for each neuron in each network-state of each recording and 
the average firing rate of the same neuron in the same network state. 
There is no evident relation between the average firing of a neuron in a network state and 
the network properties that we computed, 
showing that the various connectivity
styles are not simply related to the 
neurons' activity.}
\label{fig:S4}
\end{figure}

\begin{figure}[htb]
\centerline{\includegraphics[width=\textwidth]{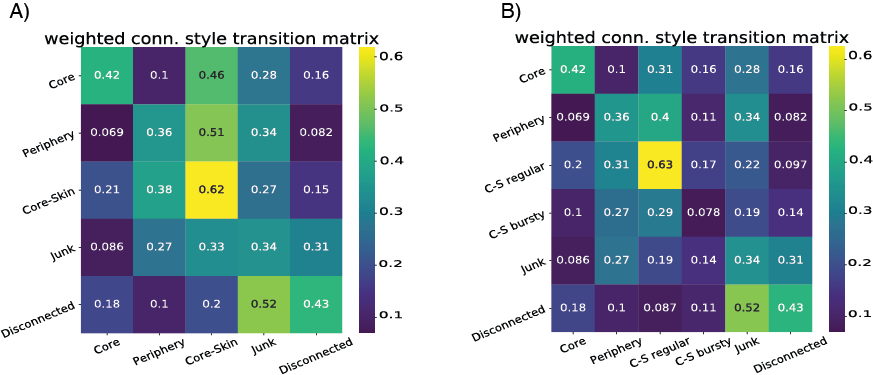}}
\caption{\textbf{A)} Transition matrix $T_{ij}$ between connectivity styles of a neuron in successive network states taking into account the core, periphery, core-skin and junk connectivity styles, as well as network states in which the neurons are not connected to the rest of the network. \textbf{B)} Transition matrix $T_{ij}$ between connectivity styles of a neuron in successive network states, taking into account the core, periphery, core-skin regular, core-skin bursty and junk connectivity styles, as well as the disconnected state of neurons. 
For both matrices, each diagonal element $T(i,i)$  represents the persistency rate of the corresponding connectivity style, i.e., 
the probability of a neuron to exhibit the same connectivity style in two successive global
network states. The non-diagonal 
matrix elements
are normalized on each row so that $\sum_{j\neq i}T(i,j)=1$.}
\label{fig:S5}
\end{figure}

\begin{figure}[htb]
\centerline{\includegraphics[width=\textwidth]{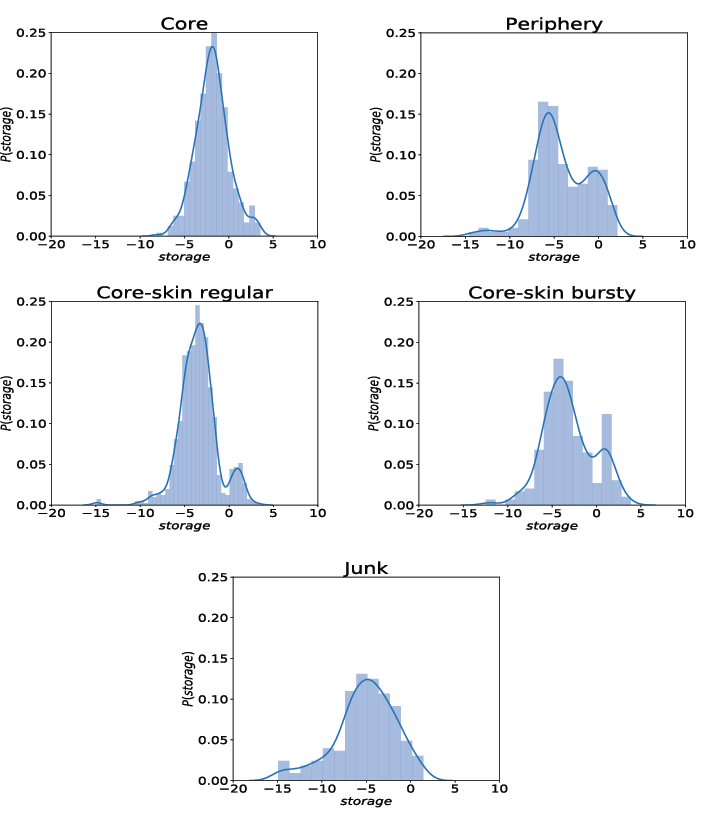}}
\caption{Density plots of the logarithm of the network-state-aggregated storage of connectivity profiles of each of the five (junk included) connectivity styles.}
\label{fig:S6}
\end{figure}

\begin{figure}[htb]
\centerline{\includegraphics[width=\textwidth]{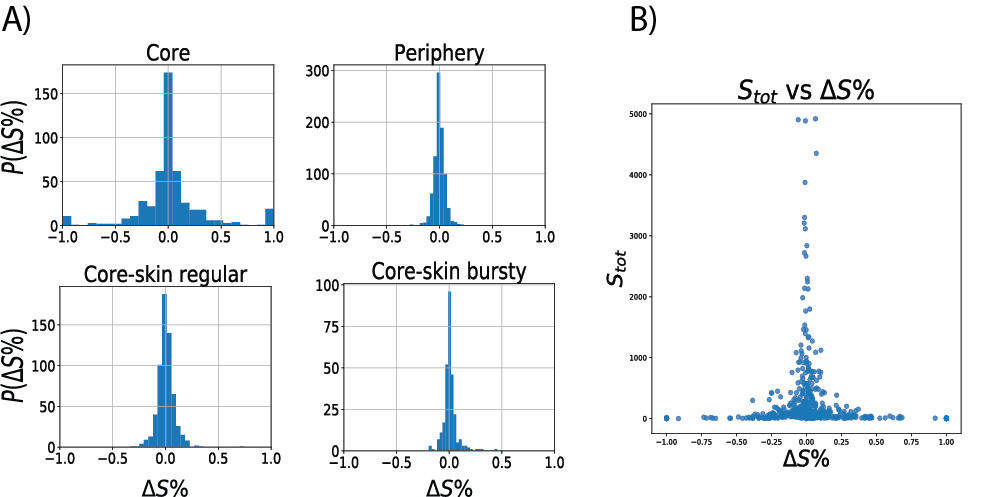}}
\caption{\textbf{A)} Distribution of values of $\Delta S\%$ for the connectivity profiles of each connectivity style. All the distributions of 
are peaked around $0$,
with the extreme values ($+1$ and $-1$)
reached only 
for \emph{core} connectivity profiles. 
The corresponding population of perfect senders and receivers represents 
a small portion of the overall number of connectivity profiles. 
\textbf{B)}
Scatterplot of the values of 
$S_{tot}$ vs. $\Delta S\%$, the former representing the total aggregate strength of a neuron 
during a state: $S_{tot}^{i,h}=s^{i,h}_{in}+s^{i,h}_{out}$. 
The scatterplot shows that perfect sender/receiver neurons ($\Delta S\%=\pm1$) correspond to very 
low values of $S_{tot}$, hence they correspond to
very weakly connected neurons.}
\label{fig:S7}
\end{figure}

\end{document}